\documentclass{eptcs}
\providecommand{\event}{MSFP 2018} 
\usepackage{breakurl}             
\usepackage{underscore}           
\usepackage{amsmath}
\usepackage{amssymb}
\usepackage{verbatim}

\usepackage{agda}

\DeclareUnicodeCharacter{"2115}{\ensuremath{\mathbb{N}}}
\DeclareUnicodeCharacter{"2192}{\ensuremath{\rightarrow}}
\DeclareUnicodeCharacter{"27E6}{\ensuremath{[\![}}
\DeclareUnicodeCharacter{"27E7}{\ensuremath{]\!]}}
\DeclareUnicodeCharacter{"2090}{\ensuremath{_a}}
\DeclareUnicodeCharacter{"AC}{\ensuremath{\neg}}
\DeclareUnicodeCharacter{"22A5}{\ensuremath{\bot}}
\DeclareUnicodeCharacter{"2081}{\ensuremath{_1}}
\DeclareUnicodeCharacter{"222A}{\ensuremath{\cup}}
\DeclareUnicodeCharacter{"21D2}{\ensuremath{\Rightarrow}}
\DeclareUnicodeCharacter{"2082}{\ensuremath{_2}}
\DeclareUnicodeCharacter{"228E}{\ensuremath{\uplus}}
\DeclareUnicodeCharacter{"D7}{\ensuremath{\times}}
\DeclareUnicodeCharacter{"2237}{\ensuremath{:\!:}}
\DeclareUnicodeCharacter{"2194}{\ensuremath{\leftrightarrow}}
\DeclareUnicodeCharacter{"2218}{\ensuremath{\circ}}
\DeclareUnicodeCharacter{"2200}{\ensuremath{\forall}}
\DeclareUnicodeCharacter{"2203}{\ensuremath{\exists}}
\DeclareUnicodeCharacter{"2032}{\ensuremath{^\prime}}
\DeclareUnicodeCharacter{"2080}{\ensuremath{_0}}
\DeclareUnicodeCharacter{"2205}{\ensuremath{\varnothing}}
\DeclareUnicodeCharacter{"2261}{\ensuremath{\equiv}}
\DeclareUnicodeCharacter{"2228}{\ensuremath{\vee}}
\DeclareUnicodeCharacter{"3C6}{\ensuremath{\phi}}
\DeclareUnicodeCharacter{"3A3}{\ensuremath{\Sigma}}
\DeclareUnicodeCharacter{"3BB}{\ensuremath{\lambda}}
\DeclareUnicodeCharacter{"3C8}{\ensuremath{\psi}}
\DeclareUnicodeCharacter{"3A0}{\ensuremath{\Pi}}

\newcommand{\nope}[0]{\textasciitilde}
\newcommand{\impl}[0]{\Rightarrow}

\newcommand{\cons}[1]{\AgdaInductiveConstructor{#1}}
\newcommand{\field}[1]{\AgdaField{#1}}
\newcommand{\func}[1]{\AgdaFunction{#1}}
\newcommand{\data}[1]{\AgdaDatatype{#1}}
\newcommand{\asym}[1]{\AgdaSymbol{#1}}
\newcommand{\bound}[1]{\AgdaBound{#1}}
\newcommand{\record}[1]{\AgdaRecord{#1}}
\newcommand{\num}[1]{\AgdaNumber{#1}}

\title{Formalizing Constructive Quantifier Elimination in Agda}
\author{Jeremy Pope
\institute{University of Gothenburg\\ Gothenburg, Sweden}
\email{guspopje@student.gu.se}
}
\def\titlerunning{Formalizing Constructive Quantifier Elimination in Agda}
\def\authorrunning{Jeremy Pope}

\begin{document}
\maketitle

\begin{abstract}
In this paper a constructive formalization of quantifier elimination is presented, based on a classical formalization by Tobias Nipkow.
The formalization is implemented and verified in the programming language/proof assistant Agda.
It is shown that, as in the classical case, the ability to eliminate a single existential quantifier may be generalized to full quantifier elimination and consequently a decision procedure.
The latter is shown to have strong properties under a  constructive metatheory, such as the generation of witnesses and counterexamples.
Finally, this is demonstrated on a minimal theory on the natural numbers.
\end{abstract}

\section{Introduction}

\subsection{Predicate Logic and Quantifier Elimination}

A proposition in predicate logic is formed in one of three ways: from an atom, by linking propositions together using a connective (such as $\vee$ or $\impl$), or by quantifying a proposition with $\forall$ or $\exists$.
Neglecting the internal structure of atoms, it is the third that sets predicate logic apart from propositional logic; the quantifiers greatly enhance the expressiveness of the language.

A drawback of predicate logic is that the truth of a proposition is no longer easy to determine.
With propositional logic an exhaustive enumeration is possible, but this is not so in predicate logic: to do so on a proposition such as $\forall x. (x \ne x+1)$ would require verifying $x \ne x + 1$ for every possible value of $x$, which---depending on our choice of domain---could be infinite.

There is not always a way around this; predicate logic is indeed undecidable in the general case.
However, a number of specific theories within predicate logic are in fact decidable---and not simply by admitting exhaustive enumeration.
Rather, decidability is shown through \emph{quantifier elimination}.

The idea behind quantifier elimination is to devise a method to transform any given proposition into an equivalent one without quantifiers.
The latter can typically be decided very easily, and by virtue of the equivalence the decision applies to the original proposition as well.
This allows any proposition in the theory to be decided, rendering the theory decidable.

\subsection{Classical and Constructive Logic}\label{subsec:constructive}

If quantifier elimination is proven to be possible for a theory, that proof is in turn carried out in another theory, referred to as the \emph{metatheory}.

Both the theory and metatheory can vary with respect to being classical or constructive.
A classical (meta)theory, by virtue of the law of excluded middle, allows the use of quantifier dualities, De Morgan's laws, proof by contradiction, and other results of classical logic.

In a constructive (meta)theory, however, the above are no longer a priori available.
Despite such constraints, there are several advantages.
One of the most significant---beside any philosophical arguments for constructivism---is that a proof of existence requires a \emph{witness}, an actual value that meets the specified criteria.
For example, a constructive proof of $\exists x.x > 2$ would necessarily consist of a value for $x$, and a proof that it is greater than two.
As a consequence, not only does a constructive decision procedure determine whether a proposition is true or not, but it provides significantly more information about \emph{how}.

\subsection{Formalization}

The difficulty of quantifier elimination depends on the theory in question, but even for simple theories it is quite high---great care must be taken to ensure that the method is sound.
Moreover, applying the procedure to a complicated proposition is likely impractical for a human (especially if it involves conversion to disjunctive normal form, which can result in a large growth in the number of terms).
These factors encourage computer formalization of quantifier elimination---both in implementing the procedures, and verifying that they are correct.

Implementation by itself is conceptually straightforward: propositions are represented by some datatype, and quantifier elimination as procedure(s) that manipulate objects of that datatype.
Verification makes matters more complicated; the implementation must be accompanied by a proof of its correctness, which certifies that the quantifier elimination procedure always produces a proposition that is equivalent to the input (and quantifier-free).
To facilitate this, both the implementation and correctness proof are typically written in a proof assistant.

\subsection{History, and Related Work}\label{subsec:hist}

The technique of quantifier elimination has been used over the last century to prove the decidability of a number of theories. Notable examples include: real and algebraically closed fields, by Alfred Tarski \cite{atdt}; several theories on the natural numbers under a constructive metatheory, by Jacques Herbrand \cite{herbrand}; and Presburger arithmetic (addition on the natural numbers), by Moj\.{z}esz Presburger \cite{presbryan}.

Following G\"{o}del's incompleteness theorems~\cite{ginc} came several negative results in decidability.
One such result is the essential undecidability of Robinson arithmetic \cite{robinson} (addition and multiplication on the natural numbers), which effectively rules out the possibility of a decision procedure for general arithmetic.

More recently, several (positive) results have been revisited in the context of computer-verified proofs. Examples include Tobias Nipkow's framework for and application of quantifier elimination in Isabelle \cite{nipkowjar}; Assia Mahboubi and Cyril Cohen's Coq formalizations of the decidability of real \cite{mahboubircf} and algebraically \cite{mahboubiacf} closed fields; and Guillaume Allais' constructive proof of the decidability of Presburger arithmetic in Agda \cite{allaisgit}.

This paper presents a general framework somewhat similar to that of Nipkow~\cite{nipkowjar}, however carried out in Agda under a constructive metatheory.\footnote{And, alas, without reflection for the time being.}
It is applied to one of the theories Herbrand~\cite{herbrand} showed to be decidable, using a technique similar to the one that he presented.
It also bears substantial similarity to Allais's Presburger solver~\cite{allaisgit}, despite being developed largely independently: correspondance between this author and Allais, during which the latter generously shared the source code of his (at the time unpublished) work, only occurred after the majority of the work presented in this paper had been completed.
Certain presentational aspects were influenced by Allais' work, however.
More importantly, several apparent novelties of this work---such as the trick used to obviate the need for prenex form---in fact appeared earlier in the former.
The primary contribution of this work is therefore its relative generality.

\subsection{Organization}

The remainder of this paper is organized as follows:
First, brief background information is given on several theoretical aspects (Section~\ref{sec:bkg}).
Next, a theory-independent formalization of quantifier elimination is shown (Section~\ref{sec:qeqe}),
followed by an application to a theory on the natural numbers (Sections \ref{sec:suc} and \ref{sec:demon}).
Finally, possibilities for further development are discussed (Section~\ref{sec:remarks}).

The source code for the project (excluding the Agda standard library) is available on GitHub.\footnote{https://github.com/guspopje/agda-qelim} At the time of this paper's writing, the code compiles with Agda version 2.5.2, and version 0.13 of the standard library.

\section{Theoretical Background}\label{sec:bkg}

\subsection{Quantifier Elimination}\label{subsec:bkgqe}

Rather than attempting to remove all quantifiers at once, an incremental approach is usually taken, dramatically reducing the scope of the problem.
A procedure is devised to remove a single quantifier, often $\exists$, from an otherwise quantifier-free proposition:
$$\exists x.\phi \iff \psi,$$
where $\phi$ and $\psi$ are quantifier-free.
Using the quantifier duality $\forall x.\phi \iff \neg \exists x. \neg \phi$ (in a classical theory) this can be adapted to remove $\forall$ as well.
If the full proposition in question (which may contain many quantifiers) is placed into \emph{prenex form}, where all of its quantifiers are pushed as far out as possible, then repeated application of the single-step procedure can clearly be used to eliminate all quantifiers from the ``inside out'':
$$\exists z.\forall y.\exists x.\phi \iff \exists z.\forall y.\rho \iff \exists z.\sigma \iff \psi,$$
noting that $\phi$, $\rho$, $\sigma$, and $\psi$ are all quantifier-free.
The same recursive strategy can just as well be used without placing the proposition into prenex form.\footnote{This is actually quite important; transforming a proposition into prenex form uses quantifier dualities that are not valid in constructive logic.}

To narrow the problem even further, the quantifier-free sub-proposition $\phi$ can be placed into disjunctive normal form (DNF):
$$\phi \iff C_1 \vee C_2 \vee \ldots \vee C_n$$
where each $C_i$ is a conjunction of literals (a literal being an atomic formula or its negation).
This is useful because existential quantification distributes across disjunction:
$$\exists x.\phi \iff \exists x. (C_1 \vee C_2 \vee \ldots \vee C_n) \iff (\exists x. C_1) \vee (\exists x. C_2) \vee \ldots \vee (\exists x. C_n).$$
As a result, elimination can be carried out on each conjunction separately, reducing the problem to quantifier elimination on conjunctions of literals.

Once a quantifier elimination procedure has been shown, decidability of the theory is obtained---provided that all quantifier-free propositions are decidable.
The latter requirement is trivially true for theories where atomic formulae represent decidable relations (e.g. equality on the natural numbers).

\subsection{Agda}

Agda~\cite{agda}, the programming language/proof assistant used in this paper, is based on intuitionistic type theory.
As a result it is constructive, with the consequences described in Section~\ref{subsec:constructive}.
Further information about the language is available from Agda's website and the various tutorials listed there~\cite{agdatut}.

\subsection{Theory, Metatheory, and Semantics}\label{subsec:bkgtms}

In quantifier elimination, and proofs about logic systems in general, there are frequently two layers: the theory $T$ under consideration, expressed in the \emph{object language}, and the metatheory $M$ in which $T$ is analyzed, expressed in the \emph{metalanguage}.

The notions of equivalance and decidability (as related to quantifier elimination) necessitate that a notion of provability or truth be associated with $T$.
One option is to define a proof system directly for $T$, as in Herbrand's thesis~\cite{herbrand}.
This allows a syntactic treatment, notions such as ``equivalent in $T$'' and ``provable in $T$'', and consequently strong separation between theory and metatheory.\footnote{In Herbrand's case, this allows the analysis of a classical theory under a constructive metatheory.}

Another option, as taken by Tarski~\cite{tarski44}, Nipkow~\cite{nipkowjar}, and this project, is to instead define the semantics of propositions of $T$, as propositions in $M$:
$$[\![\,\cdot\,]\!] : T \to M.$$
This is typically accomplished recursively, mapping each each connective or quantifier in $T$ to the corresponding one in $M$.
In the case of this paper, $T$ is an arbitrary theory of first order logic (subject to minor constraints), and $M$ is the flavor of Intuitionistic Type Theory used by Agda.
A notable consequence of $M$ being constructive (and the definition of $[\![\,\cdot\,]\!]$ used in this paper) is that the semantics of $T$ are constructive as well.

With this approach, quantifier elimination produces a proposition that is \emph{semantically} equivalent to the original, and in the end it is the semantics of $T$ that are proven to be decidable (as opposed to $T$ itself, which is not possible without a proof system of its own).
As the semantics lie in $M$, this means that decidability is shown for a fragment of $M$.

\subsection{De Bruijn Indices}\label{subsec:debruijn2}

One of the difficulties in formalizing a theory is the handling of free and bound variables.
With a traditional ``named variable'' approach, extra conditions must be added to prevent substitutions that would capture free variables.\footnote{An example from lambda calculus is the (invalid) beta reduction of $(\lambda x.\lambda y.x)y$ to $\lambda y.y$.}

An alternative is to use De Bruijn indices, where each occurrence of a bound variable is denoted instead by a number that indicates how many variable-binders ``deep'' the occurrence is from the binder to which it refers.
For example, the proposition
\[\forall x.(x \le 4 \vee (\exists y.x = y + 5))\]
is represented as
\[\forall.(\framebox{0} \le 4 \vee (\exists. \framebox{1} = \framebox{0} + 5)).\]
Here it is noted that within the scope of the \(\exists\), the index \(\framebox{0}\) refers to the \(\exists\) (i.e., the variable \(y\)), and \(\framebox{1}\) refers to the quantifier one layer out, namely \(\forall\) (i.e., the variable \(x\)).

Free variables can be treated the same way; their indices simply point outside of the visible formula (for example, the rightmost variable in \(\exists.(\framebox{0} \ge \framebox{3})\)).
The benefits include dramatically simpler rules for substitution, alpha equivalence for free, and ease in associating values with free variables.

With named variables, the latter (referred to as an \emph{environment}) is accomplished via a mapping from names to values, while with de Bruijn indices only a list of values is required---the list corresponds to the ``missing'' layers of quantifiers to which the free variables refer.
For example, the proposition \(\forall.\exists.(\framebox{4} \ge \framebox{7})\) requires a list with at least six values.

In this paper, such a bound is referred to as the \emph{arity} of a proposition, arising from the interpretation of a proposition as a function of its free variables.

\section{Theory-Independent Work}\label{sec:qeqe}

\subsection{Atoms}\label{subsec:atm}

In the interest of generality, atomic formulae are not represented by a fixed type, but by a type given as a \emph{module parameter}.
The type is indexed by a natural numbers $n$ representing its arity (an upper bound on its free variables, as discussed in Section~\ref{subsec:debruijn2}):
\begin{code}
\>[6]\AgdaField{Atom} \AgdaSymbol{:} \AgdaDatatype{ℕ} \AgdaSymbol{→} \AgdaPrimitiveType{Set}\<%
\end{code}
The internal structure of an \field{Atom} is completely unspecified.

The semantics of \field{Atom} is also given by way of module parameters.
First, the set of values which variables may take:

\begin{code}
\>[6]\AgdaField{Y} \AgdaSymbol{:} \AgdaPrimitiveType{Set}\<%
\end{code}
Then, a function which gives the semantics for an atom:

\begin{code}
\>[6]\AgdaField{⟦\_⟧ₐ} \AgdaSymbol{:} \AgdaSymbol{\{}\AgdaBound{n} \AgdaSymbol{:} \AgdaDatatype{ℕ}\AgdaSymbol{\}} \AgdaSymbol{→} \AgdaField{Atom} \AgdaBound{n} \AgdaSymbol{→} \AgdaDatatype{Vec} \AgdaField{Y} \AgdaBound{n} \AgdaSymbol{→} \AgdaPrimitiveType{Set}\<%
\end{code}
The implicit parameter \bound{n} \asym{:} \data{ℕ} is the arity of the atom, and the following parameter of type \data{Atom} \bound{n} is the atom itself.
The last parameter, of type \data{Vec} \field{Y} \bound{n}, is the \emph{environment}: a list (vector) of length \bound{n} of values for the free variables in the atom (see Section~\ref{subsec:debruijn2}).
This, in a sense, forces \field{Atom} to use de Bruijn indices internally---no names are associated with the values in the enviroment.
Moreover, since the enviroment for an \field{Atom} \bound{n} is a list of \bound{n} values, the effective arity of the atom is restricted to \bound{n}, as intended.

Additionally, it is required that the semantics of \field{Atom} be decidable under any given enviroment.
This is often the case (as discussed in Section~\ref{subsec:bkgqe}, and is equivalent to the semantics of all quantifier-free propositions being decidable.
As it turns out, for a constructive theory this is important not only for decidability but for quantifier elimination itself, as will be seen later on.
Another module parameter is used to implement this requirement:
\begin{code}
\>[6]\AgdaField{⟦\_⟧ₐ?} \AgdaSymbol{:} \AgdaSymbol{\{}\AgdaBound{n} \AgdaSymbol{:} \AgdaDatatype{ℕ}\AgdaSymbol{\}} \AgdaSymbol{(}\AgdaBound{a} \AgdaSymbol{:} \AgdaField{Atom} \AgdaBound{n}\AgdaSymbol{)} \AgdaSymbol{(}\AgdaBound{e} \AgdaSymbol{:} \AgdaDatatype{Vec} \AgdaField{Y} \AgdaBound{n}\AgdaSymbol{)} \AgdaSymbol{→} \AgdaDatatype{Dec} \AgdaSymbol{(}\AgdaField{⟦} \AgdaBound{a} \AgdaField{⟧ₐ} \AgdaBound{e}\AgdaSymbol{)}\<%
\end{code}
The \data{Dec} type family, from Agda's standard library, is indexed by a type (\bound{A}).
An object of type \data{Dec} \bound{A} is a decision for \bound{A}: either a proof that \bound{A} is inhabited (\cons{yes} \bound{a}, where \bound{a} \asym{:} \bound{A}), or a proof that it is not (\cons{no} \bound{x}, where \bound{x} \asym{:} \func{¬} \bound{A}, i.e. \bound{x} \asym{:} \bound{A} \asym{→} \asym{⊥}).

For organizational purposes the above are grouped into a record type, forming an abstract representation of atoms with decidable semantics:
\begin{code}%
\>[0]\AgdaIndent{2}{}\<[2]%
\>[2]\AgdaKeyword{record} \AgdaRecord{DecAtom} \AgdaSymbol{:} \AgdaPrimitiveType{Set₁} \AgdaKeyword{where}\<%
\\
\>[2]\AgdaIndent{4}{}\<[4]%
\>[4]\AgdaKeyword{field}\<%
\\
\>[4]\AgdaIndent{6}{}\<[6]%
\>[6]\AgdaField{Atom} \AgdaSymbol{:} \AgdaDatatype{ℕ} \AgdaSymbol{→} \AgdaPrimitiveType{Set}\<%
\\
\>[4]\AgdaIndent{6}{}\<[6]%
\>[6]\AgdaField{Y} \AgdaSymbol{:} \AgdaPrimitiveType{Set}\<%
\\
\>[4]\AgdaIndent{6}{}\<[6]%
\>[6]\AgdaField{⟦\_⟧ₐ} \AgdaSymbol{:} \AgdaSymbol{\{}\AgdaBound{n} \AgdaSymbol{:} \AgdaDatatype{ℕ}\AgdaSymbol{\}} \AgdaSymbol{→} \AgdaField{Atom} \AgdaBound{n} \AgdaSymbol{→} \AgdaDatatype{Vec} \AgdaField{Y} \AgdaBound{n} \AgdaSymbol{→} \AgdaPrimitiveType{Set}\<%
\\
\>[4]\AgdaIndent{6}{}\<[6]%
\>[6]\AgdaField{⟦\_⟧ₐ?} \AgdaSymbol{:} \AgdaSymbol{\{}\AgdaBound{n} \AgdaSymbol{:} \AgdaDatatype{ℕ}\AgdaSymbol{\}} \AgdaSymbol{(}\AgdaBound{a} \AgdaSymbol{:} \AgdaField{Atom} \AgdaBound{n}\AgdaSymbol{)} \AgdaSymbol{(}\AgdaBound{e} \AgdaSymbol{:} \AgdaDatatype{Vec} \AgdaField{Y} \AgdaBound{n}\AgdaSymbol{)} \AgdaSymbol{→} \AgdaDatatype{Dec} \AgdaSymbol{(}\AgdaField{⟦} \AgdaBound{a} \AgdaField{⟧ₐ} \AgdaBound{e}\AgdaSymbol{)}\<%
\\
\>\<%
\end{code}
A single module parameter of type \record{DecAtom} is used in lieu of four separate parameters.

\subsection{Representation of Propositions}

Propositions are represented by following datatype \data{Prop}.
Its constructors allow the formation of a proposition from an atom, or from other propositions by way of the typical connectives and quantifiers.
\begin{code}%
\>[2]\AgdaIndent{4}{}\<[4]%
\>[4]\AgdaKeyword{data} \AgdaDatatype{Prop} \AgdaSymbol{(}\AgdaBound{n} \AgdaSymbol{:} \AgdaDatatype{ℕ}\AgdaSymbol{)} \AgdaSymbol{:} \AgdaPrimitiveType{Set} \AgdaKeyword{where}\<%
\\
\>[4]\AgdaIndent{6}{}\<[6]%
\>[6]\AgdaInductiveConstructor{atom} \AgdaSymbol{:} \AgdaField{Atom} \AgdaBound{n} \AgdaSymbol{→} \AgdaDatatype{Prop} \AgdaBound{n}\<%
\\
\>[4]\AgdaIndent{6}{}\<[6]%
\>[6]\AgdaInductiveConstructor{⊥⊥} \<[11]%
\>[11]\AgdaSymbol{:} \AgdaDatatype{Prop} \AgdaBound{n}\<%
\\
\>[4]\AgdaIndent{6}{}\<[6]%
\>[6]\AgdaInductiveConstructor{\_∨\_} \<[11]%
\>[11]\AgdaSymbol{:} \AgdaDatatype{Prop} \AgdaBound{n} \AgdaSymbol{→} \AgdaDatatype{Prop} \AgdaBound{n} \AgdaSymbol{→} \AgdaDatatype{Prop} \AgdaBound{n}\<%
\\
\>[4]\AgdaIndent{6}{}\<[6]%
\>[6]\AgdaInductiveConstructor{\_∧\_} \<[11]%
\>[11]\AgdaSymbol{:} \AgdaDatatype{Prop} \AgdaBound{n} \AgdaSymbol{→} \AgdaDatatype{Prop} \AgdaBound{n} \AgdaSymbol{→} \AgdaDatatype{Prop} \AgdaBound{n}\<%
\\
\>[4]\AgdaIndent{6}{}\<[6]%
\>[6]\AgdaInductiveConstructor{\_⇒\_} \<[11]%
\>[11]\AgdaSymbol{:} \AgdaDatatype{Prop} \AgdaBound{n} \AgdaSymbol{→} \AgdaDatatype{Prop} \AgdaBound{n} \AgdaSymbol{→} \AgdaDatatype{Prop} \AgdaBound{n}\<%
\\
\>[4]\AgdaIndent{6}{}\<[6]%
\>[6]\AgdaInductiveConstructor{E\_} \<[11]%
\>[11]\AgdaSymbol{:} \AgdaDatatype{Prop} \AgdaSymbol{(}\AgdaInductiveConstructor{suc} \AgdaBound{n}\AgdaSymbol{)} \AgdaSymbol{→} \AgdaDatatype{Prop} \AgdaBound{n}\<%
\\
\>[4]\AgdaIndent{6}{}\<[6]%
\>[6]\AgdaInductiveConstructor{A\_} \<[11]%
\>[11]\AgdaSymbol{:} \AgdaDatatype{Prop} \AgdaSymbol{(}\AgdaInductiveConstructor{suc} \AgdaBound{n}\AgdaSymbol{)} \AgdaSymbol{→} \AgdaDatatype{Prop} \AgdaBound{n}\<%
\end{code}
Negation is defined for convenience:
\begin{code}%
\>[0]\AgdaIndent{4}{}\<[4]%
\>[4]\AgdaFunction{\textasciitilde{}\_} \AgdaSymbol{:} \AgdaSymbol{\{}\AgdaBound{n} \AgdaSymbol{:} \AgdaDatatype{ℕ}\AgdaSymbol{\}} \AgdaSymbol{→} \AgdaDatatype{Prop} \AgdaBound{n} \AgdaSymbol{→} \AgdaDatatype{Prop} \AgdaBound{n}\<%
\\
\>[0]\AgdaIndent{4}{}\<[4]%
\>[4]\AgdaFunction{\textasciitilde{}} \AgdaBound{φ} \AgdaSymbol{=} \AgdaBound{φ} \AgdaInductiveConstructor{⇒} \AgdaInductiveConstructor{⊥⊥}\<%
\end{code}

It is noted that because the semantics of a proposition is not a priori decidable, under a constructive (meta)theory propositions cannot be reduced to a more minimal set of connectives/quantifiers, as would be typical in a classical setting.

The quantifiers \cons{E\_} and \cons{A\_} reflect the use of de Bruijn indices (Section~\ref{subsec:debruijn2}): neither constructor accepts any indication of which variable is to be quantified (recall that with de Bruijn indices this is not needed), and both decrement the arity (by virtue of binding one of the free variables in the quantified proposition).

\subsection{Semantics of Propositions}\label{subsec:propsem}

The semantics of a proposition is then defined recursively from \field{⟦\_⟧ₐ}:
\begin{code}%
\>[0]\AgdaIndent{4}{}\<[4]%
\>[4]\AgdaFunction{⟦\_⟧} \AgdaSymbol{:} \AgdaSymbol{\{}\AgdaBound{n} \AgdaSymbol{:} \AgdaDatatype{ℕ}\AgdaSymbol{\}} \AgdaSymbol{→} \AgdaDatatype{Prop} \AgdaBound{n} \AgdaSymbol{→} \AgdaDatatype{Vec} \AgdaField{Y} \AgdaBound{n} \AgdaSymbol{→} \AgdaPrimitiveType{Set}\<%
\\
\>[0]\AgdaIndent{4}{}\<[4]%
\>[4]\AgdaFunction{⟦} \AgdaInductiveConstructor{⊥⊥} \AgdaFunction{⟧} \<[16]%
\>[16]\AgdaBound{ys} \AgdaSymbol{=} \AgdaDatatype{⊥}\<%
\\
\>[0]\AgdaIndent{4}{}\<[4]%
\>[4]\AgdaFunction{⟦} \AgdaInductiveConstructor{atom} \AgdaBound{a} \AgdaFunction{⟧} \<[16]%
\>[16]\AgdaBound{ys} \AgdaSymbol{=} \AgdaField{⟦} \AgdaBound{a} \AgdaField{⟧ₐ} \AgdaBound{ys}\<%
\\
\>[0]\AgdaIndent{4}{}\<[4]%
\>[4]\AgdaFunction{⟦} \AgdaBound{φ₁} \AgdaInductiveConstructor{∨} \AgdaBound{φ₂} \AgdaFunction{⟧} \<[16]%
\>[16]\AgdaBound{ys} \AgdaSymbol{=} \AgdaSymbol{(}\AgdaFunction{⟦} \AgdaBound{φ₁} \AgdaFunction{⟧} \AgdaBound{ys}\AgdaSymbol{)} \AgdaDatatype{⊎} \AgdaSymbol{(}\AgdaFunction{⟦} \AgdaBound{φ₂} \AgdaFunction{⟧} \AgdaBound{ys}\AgdaSymbol{)}\<%
\\
\>[0]\AgdaIndent{4}{}\<[4]%
\>[4]\AgdaFunction{⟦} \AgdaBound{φ₁} \AgdaInductiveConstructor{∧} \AgdaBound{φ₂} \AgdaFunction{⟧} \<[16]%
\>[16]\AgdaBound{ys} \AgdaSymbol{=} \AgdaSymbol{(}\AgdaFunction{⟦} \AgdaBound{φ₁} \AgdaFunction{⟧} \AgdaBound{ys}\AgdaSymbol{)} \AgdaFunction{×} \AgdaSymbol{(}\AgdaFunction{⟦} \AgdaBound{φ₂} \AgdaFunction{⟧} \AgdaBound{ys}\AgdaSymbol{)}\<%
\\
\>[0]\AgdaIndent{4}{}\<[4]%
\>[4]\AgdaFunction{⟦} \AgdaBound{φ₁} \AgdaInductiveConstructor{⇒} \AgdaBound{φ₂} \AgdaFunction{⟧} \<[16]%
\>[16]\AgdaBound{ys} \AgdaSymbol{=} \AgdaSymbol{(}\AgdaFunction{⟦} \AgdaBound{φ₁} \AgdaFunction{⟧} \AgdaBound{ys}\AgdaSymbol{)} \AgdaSymbol{→} \AgdaSymbol{(}\AgdaFunction{⟦} \AgdaBound{φ₂} \AgdaFunction{⟧} \AgdaBound{ys}\AgdaSymbol{)}\<%
\\
\>[0]\AgdaIndent{4}{}\<[4]%
\>[4]\AgdaFunction{⟦} \AgdaInductiveConstructor{E} \AgdaBound{φ} \AgdaFunction{⟧} \<[16]%
\>[16]\AgdaBound{ys} \AgdaSymbol{=} \AgdaRecord{Σ} \AgdaField{Y} \AgdaSymbol{(λ} \AgdaBound{y} \AgdaSymbol{→} \AgdaFunction{⟦} \AgdaBound{φ} \AgdaFunction{⟧} \AgdaSymbol{(}\AgdaBound{y} \AgdaInductiveConstructor{∷} \AgdaBound{ys}\AgdaSymbol{))}\<%
\\
\>[0]\AgdaIndent{4}{}\<[4]%
\>[4]\AgdaFunction{⟦} \AgdaInductiveConstructor{A} \AgdaBound{φ} \AgdaFunction{⟧} \<[16]%
\>[16]\AgdaBound{ys} \AgdaSymbol{=} \AgdaSymbol{(}\AgdaBound{y} \AgdaSymbol{:} \AgdaField{Y}\AgdaSymbol{)} \AgdaSymbol{→} \AgdaSymbol{(}\AgdaFunction{⟦} \AgdaBound{φ} \AgdaFunction{⟧} \AgdaSymbol{(}\AgdaBound{y} \AgdaInductiveConstructor{∷} \AgdaBound{ys}\AgdaSymbol{))}\<%
\end{code}

Absurdity, disjunction, conjunction, and implication are respectively mapped to the empty, disjoint union, cartesian product, and function types.

The semantics of existential quantification is represented using a $\Sigma$ (dependent sum/pair) type.
Members of the resulting type are pairs consisting of a value \bound{y} \asym{:} \field{Y} and an element of the inner proposition's semantics with \bound{y} prepended to the environment, i.e., proof that the inner proposition is true with the first free variable ``set to \bound{y}''.

The semantics of universal quantification is defined similarly, but using a (dependent) function type\footnote{A $\Pi$ type, though Agda's syntax makes it of little use to write it as such.} in place of the $\Sigma$ type---\emph{all} values for \bound{y} must result in the inner proposition being true.

\subsection{Quantifier-Free Propositions}\label{subsec:qfree}

As quantifier-free propositions are of importance, a representation of this quality is defined:

\begin{code}%
\>[0]\AgdaIndent{4}{}\<[4]%
\>[4]\AgdaKeyword{data} \AgdaDatatype{QFree} \AgdaSymbol{\{}\AgdaBound{n} \AgdaSymbol{:} \AgdaDatatype{ℕ}\AgdaSymbol{\}} \AgdaSymbol{:} \AgdaDatatype{Prop} \AgdaBound{n} \AgdaSymbol{→} \AgdaPrimitiveType{Set} \AgdaKeyword{where}\<%
\\
\>[4]\AgdaIndent{6}{}\<[6]%
\>[6]\AgdaInductiveConstructor{⊥⊥} \<[10]%
\>[10]\AgdaSymbol{:} \AgdaDatatype{QFree} \AgdaInductiveConstructor{⊥⊥}\<%
\\
\>[4]\AgdaIndent{6}{}\<[6]%
\>[6]\AgdaInductiveConstructor{atom} \AgdaSymbol{:} \AgdaSymbol{(}\AgdaBound{a} \AgdaSymbol{:} \AgdaField{Atom} \AgdaBound{n}\AgdaSymbol{)} \AgdaSymbol{→} \AgdaDatatype{QFree} \AgdaSymbol{(}\AgdaInductiveConstructor{atom} \AgdaBound{a}\AgdaSymbol{)}\<%
\\
\>[4]\AgdaIndent{6}{}\<[6]%
\>[6]\AgdaInductiveConstructor{\_∨\_} \AgdaSymbol{:} \<[13]%
\>[13]\AgdaSymbol{\{}\AgdaBound{φ₁} \AgdaBound{φ₂} \AgdaSymbol{:} \AgdaDatatype{Prop} \AgdaBound{n}\AgdaSymbol{\}} \AgdaSymbol{→} \AgdaDatatype{QFree} \AgdaBound{φ₁} \AgdaSymbol{→} \AgdaDatatype{QFree} \AgdaBound{φ₂} \AgdaSymbol{→} \AgdaDatatype{QFree} \AgdaSymbol{(}\AgdaBound{φ₁} \AgdaInductiveConstructor{∨} \AgdaBound{φ₂}\AgdaSymbol{)}\<%
\\
\>[4]\AgdaIndent{6}{}\<[6]%
\>[6]\AgdaInductiveConstructor{\_∧\_} \AgdaSymbol{:} \<[13]%
\>[13]\AgdaSymbol{\{}\AgdaBound{φ₁} \AgdaBound{φ₂} \AgdaSymbol{:} \AgdaDatatype{Prop} \AgdaBound{n}\AgdaSymbol{\}} \AgdaSymbol{→} \AgdaDatatype{QFree} \AgdaBound{φ₁} \AgdaSymbol{→} \AgdaDatatype{QFree} \AgdaBound{φ₂} \AgdaSymbol{→} \AgdaDatatype{QFree} \AgdaSymbol{(}\AgdaBound{φ₁} \AgdaInductiveConstructor{∧} \AgdaBound{φ₂}\AgdaSymbol{)}\<%
\\
\>[4]\AgdaIndent{6}{}\<[6]%
\>[6]\AgdaInductiveConstructor{\_⇒\_} \AgdaSymbol{:} \<[13]%
\>[13]\AgdaSymbol{\{}\AgdaBound{φ₁} \AgdaBound{φ₂} \AgdaSymbol{:} \AgdaDatatype{Prop} \AgdaBound{n}\AgdaSymbol{\}} \AgdaSymbol{→} \AgdaDatatype{QFree} \AgdaBound{φ₁} \AgdaSymbol{→} \AgdaDatatype{QFree} \AgdaBound{φ₂} \AgdaSymbol{→} \AgdaDatatype{QFree} \AgdaSymbol{(}\AgdaBound{φ₁} \AgdaInductiveConstructor{⇒} \AgdaBound{φ₂}\AgdaSymbol{)}\<%
\\
\\
\>[0]\AgdaIndent{4}{}\<[4]%
\>[4]\AgdaFunction{\textasciitilde{}-qf\_} \AgdaSymbol{:} \AgdaSymbol{\{}\AgdaBound{n} \AgdaSymbol{:} \AgdaDatatype{ℕ}\AgdaSymbol{\}} \AgdaSymbol{\{}\AgdaBound{φ} \AgdaSymbol{:} \AgdaDatatype{Prop} \AgdaBound{n}\AgdaSymbol{\}} \AgdaSymbol{→} \AgdaDatatype{QFree} \AgdaBound{φ} \AgdaSymbol{→} \AgdaDatatype{QFree} \AgdaSymbol{(}\AgdaFunction{\textasciitilde{}} \AgdaBound{φ}\AgdaSymbol{)}\<%
\\
\>[0]\AgdaIndent{4}{}\<[4]%
\>[4]\AgdaFunction{\textasciitilde{}-qf} \AgdaBound{qf} \AgdaSymbol{=} \AgdaBound{qf} \AgdaInductiveConstructor{⇒} \AgdaInductiveConstructor{⊥⊥}\<%
\end{code}

\data{QFree} \bound{φ} is inhabited if and only if \bound{φ} is quantifier-free.

Semantically speaking, all of the connectives preserve decidability: the result of joining two semantically decidable propositions with \cons{\_∨\_}, \cons{\_∧\_}, or \cons{\_⇒\_} is also semantically decidable.
This is shown for \cons{\_⇒\_} (with semantics \asym{→}) as follows:
\begin{code}%
\>[0]\AgdaIndent{2}{}\<[2]%
\>[2]\AgdaFunction{\_→?\_} \AgdaSymbol{:} \AgdaSymbol{\{}\AgdaBound{A} \AgdaBound{B} \AgdaSymbol{:} \AgdaPrimitiveType{Set}\AgdaSymbol{\}} \AgdaSymbol{→} \AgdaDatatype{Dec} \AgdaBound{A} \AgdaSymbol{→} \AgdaDatatype{Dec} \AgdaBound{B} \AgdaSymbol{→} \AgdaDatatype{Dec} \AgdaSymbol{(}\AgdaBound{A} \AgdaSymbol{→} \AgdaBound{B}\AgdaSymbol{)}\<%
\\
\>[0]\AgdaIndent{2}{}\<[2]%
\>[2]\AgdaSymbol{\_} \<[10]%
\>[10]\AgdaFunction{→?} \AgdaSymbol{(}\AgdaInductiveConstructor{yes} \AgdaBound{b}\AgdaSymbol{)} \<[11]%
\>[11]\AgdaSymbol{=} \AgdaInductiveConstructor{yes} \AgdaSymbol{(λ} \AgdaBound{\_} \AgdaSymbol{→} \AgdaBound{b}\AgdaSymbol{)}\<%
\\
\>[0]\AgdaIndent{2}{}\<[2]%
\>[2]\AgdaSymbol{(}\AgdaInductiveConstructor{yes} \AgdaBound{a}\AgdaSymbol{)} \<[10]%
\>[10]\AgdaFunction{→?} \AgdaSymbol{(}\AgdaInductiveConstructor{no} \AgdaBound{¬b}\AgdaSymbol{)} \<[11]%
\>[11]\AgdaSymbol{=} \AgdaInductiveConstructor{no} \AgdaSymbol{(λ} \AgdaBound{f} \AgdaSymbol{→} \AgdaBound{¬b} \AgdaSymbol{(}\AgdaBound{f} \AgdaBound{a}\AgdaSymbol{))}\<%
\\
\>[0]\AgdaIndent{2}{}\<[2]%
\>[2]\AgdaSymbol{(}\AgdaInductiveConstructor{no} \AgdaBound{¬a}\AgdaSymbol{)} \<[10]%
\>[10]\AgdaFunction{→?} \AgdaSymbol{(}\AgdaInductiveConstructor{no} \AgdaBound{¬b}\AgdaSymbol{)} \<[11]%
\>[11]\AgdaSymbol{=} \AgdaInductiveConstructor{yes} \AgdaSymbol{(λ} \AgdaBound{a} \AgdaSymbol{→} \AgdaFunction{contradiction} \AgdaBound{a} \AgdaBound{¬a}\AgdaSymbol{)}\<%
\end{code}
The same property can be shown for \cons{\_∨\_} and \cons{\_∧\_} (with semantics \data{\_⊎\_} and \data{\_×\_}) in a similar manner, resulting in the following two functions:
\begin{code}%
\>[0]\AgdaIndent{2}{}\<[2]%
\>[2]\AgdaFunction{\_×?\_} \<[10]%
\>[10]\AgdaSymbol{:} \AgdaSymbol{\{}\AgdaBound{A} \AgdaBound{B} \AgdaSymbol{:} \AgdaPrimitiveType{Set}\AgdaSymbol{\}} \AgdaSymbol{→} \AgdaDatatype{Dec} \AgdaBound{A} \AgdaSymbol{→} \AgdaDatatype{Dec} \AgdaBound{B} \AgdaSymbol{→} \AgdaDatatype{Dec} \AgdaSymbol{(}\AgdaBound{A} \AgdaFunction{×} \AgdaBound{B}\AgdaSymbol{)}\<%
\\
\>[2]\AgdaFunction{\_⊎?\_} \<[10]%
\>[10]\AgdaSymbol{:} \AgdaSymbol{\{}\AgdaBound{A} \AgdaBound{B} \AgdaSymbol{:} \AgdaPrimitiveType{Set}\AgdaSymbol{\}} \AgdaSymbol{→} \AgdaDatatype{Dec} \AgdaBound{A} \AgdaSymbol{→} \AgdaDatatype{Dec} \AgdaBound{B} \AgdaSymbol{→} \AgdaDatatype{Dec} \AgdaSymbol{(}\AgdaBound{A} \AgdaDatatype{⊎} \AgdaBound{B}\AgdaSymbol{)}\<%
\end{code}
It is also noted that the semantics of \cons{⊥⊥}, namely \data{⊥}, is trivially decidable.

Given the above and that the semantics for atoms are decidable (\field{⟦\_⟧ₐ?}), it follows by induction that the semantics of any quantifier-free proposition is decidable:
\begin{code}%
\>[0]\AgdaIndent{4}{}\<[4]%
\>[4]\AgdaFunction{qfree-dec} \AgdaSymbol{:} \AgdaSymbol{\{}\AgdaBound{n} \AgdaSymbol{:} \AgdaDatatype{ℕ}\AgdaSymbol{\}} \AgdaSymbol{→} \AgdaSymbol{(}\AgdaBound{φ} \AgdaSymbol{:} \AgdaDatatype{Prop} \AgdaBound{n}\AgdaSymbol{)} \AgdaSymbol{→} \AgdaDatatype{QFree} \AgdaBound{φ} \AgdaSymbol{→} \AgdaSymbol{(}\AgdaBound{e} \AgdaSymbol{:} \AgdaDatatype{Vec} \AgdaField{Y} \AgdaBound{n}\AgdaSymbol{)} \AgdaSymbol{→} \AgdaDatatype{Dec} \AgdaSymbol{(}\AgdaFunction{⟦} \AgdaBound{φ} \AgdaFunction{⟧} \AgdaBound{e}\AgdaSymbol{)}\<%
\end{code}

\subsection{Quantifier Elimination}\label{subsec:qeqeqe}

As discussed in Section~\ref{subsec:bkgqe}, quantifier elimination is typically accomplished by eliminating existential quantifiers one by one, from the ``inside out''.
It is performed in that order so that when a quantifier is being eliminated, the enclosed proposition is already quantifier-free, simplifying the problem significantly.

The method by which a single quantifier is eliminated depends on the theory under consideration, making it impossible to directly define (whilst maintaining generality).
Instead---in a similar manner to \record{DecAtom}---it is defined abstractly with a record type \record{QE} which captures the necessary properties of a single-step elimination procedure.
A specific implementation takes the form of an object \bound{qe} \asym{:} \record{QE}.

\begin{code}%
\>[0]\AgdaIndent{4}{}\<[4]%
\>[4]\AgdaKeyword{record} \AgdaRecord{QE} \AgdaSymbol{:} \AgdaPrimitiveType{Set} \AgdaKeyword{where}\<%
\\
\>[4]\AgdaIndent{6}{}\<[6]%
\>[6]\AgdaKeyword{field}\<%
\\
\>[6]\AgdaIndent{8}{}\<[8]%
\>[8]\AgdaField{elim} \<[14]%
\>[14]\AgdaSymbol{:} \AgdaSymbol{\{}\AgdaBound{n} \AgdaSymbol{:} \AgdaDatatype{ℕ}\AgdaSymbol{\}} \AgdaSymbol{(}\AgdaBound{φ} \AgdaSymbol{:} \AgdaDatatype{Prop} \AgdaSymbol{(}\AgdaInductiveConstructor{suc} \AgdaBound{n}\AgdaSymbol{))} \AgdaSymbol{→} \AgdaDatatype{QFree} \AgdaBound{φ} \AgdaSymbol{→} \AgdaDatatype{Prop} \AgdaBound{n}\<%
\\
\>[6]\AgdaIndent{8}{}\<[8]%
\>[8]\AgdaField{qfree} \<[14]%
\>[14]\AgdaSymbol{:} \AgdaSymbol{\{}\AgdaBound{n} \AgdaSymbol{:} \AgdaDatatype{ℕ}\AgdaSymbol{\}} \AgdaSymbol{(}\AgdaBound{φ} \AgdaSymbol{:} \AgdaDatatype{Prop} \AgdaSymbol{(}\AgdaInductiveConstructor{suc} \AgdaBound{n}\AgdaSymbol{))} \AgdaSymbol{(}\AgdaBound{qf} \AgdaSymbol{:} \AgdaDatatype{QFree} \AgdaBound{φ}\AgdaSymbol{)} \AgdaSymbol{→} \AgdaDatatype{QFree} \AgdaSymbol{(}\AgdaField{elim} \AgdaBound{φ} \AgdaBound{qf}\AgdaSymbol{)}\<%
\\
\>[6]\AgdaIndent{8}{}\<[8]%
\>[8]\AgdaField{equiv} \<[14]%
\>[14]\AgdaSymbol{:} \AgdaSymbol{\{}\AgdaBound{n} \AgdaSymbol{:} \AgdaDatatype{ℕ}\AgdaSymbol{\}} \AgdaSymbol{(}\AgdaBound{φ} \AgdaSymbol{:} \AgdaDatatype{Prop} \AgdaSymbol{(}\AgdaInductiveConstructor{suc} \AgdaBound{n}\AgdaSymbol{))} \AgdaSymbol{(}\AgdaBound{qf} \AgdaSymbol{:} \AgdaDatatype{QFree} \AgdaBound{φ}\AgdaSymbol{)} \AgdaSymbol{(}\AgdaBound{e} \AgdaSymbol{:} \AgdaDatatype{Vec} \AgdaField{Y} \AgdaBound{n}\AgdaSymbol{)} \AgdaSymbol{→}\<%
\\
\>[8]\AgdaIndent{10}{}\<[10]%
\>[10]\AgdaFunction{⟦} \AgdaInductiveConstructor{E} \AgdaBound{φ} \AgdaFunction{⟧} \AgdaBound{e} \AgdaFunction{↔} \AgdaFunction{⟦} \AgdaField{elim} \AgdaBound{φ} \AgdaBound{qf} \AgdaFunction{⟧} \AgdaBound{e}\<%
\end{code}

The field \field{elim} represents the single-step elimination procedure itself, accepting a quantifier-free proposition with up to $n+1$ free variables and producing one with up to $n$.
It is noted that the input to \field{elim} does not contain the existential quantifier to eliminate, rather it is implied---for example, to eliminate the quantifier from \cons{E} \bound{φ}, the \field{elim} procedure is invoked on just \bound{φ}.
The field \field{qfree} represents a proof that \field{elim} always produces a quantifier-free proposition. Finally, \field{equiv} establishes \field{elim}'s correctness---that the propositions \cons{E} \bound{φ} and \field{elim} \bound{φ} \bound{...} are semantically equivalent.\footnote{The notation \bound{A} \asym{↔} \bound{B} is defined as \asym{(} \bound{A} \asym{→} \bound{B} \asym{)} \asym{×} \asym{(} \bound{B} \asym{→} \bound{A} \asym{)}.}

Such a single-step procedure can then be ``lifted'' to eliminate all quantifiers from a proposition via recursion on the proposition's structure (the general approach, as stated before, being to eliminate quantifiers from the inside out).
The cases are as follows:
\begin{enumerate}
\item The absurd proposition (\cons{⊥⊥}); it is left unchanged.
\item An atom; it is left unchanged.
\item A proposition formed from disjunction, conjunction, or implication (\cons{∨}, \cons{∧}, or \cons{⇒}); the sub-proposition(s) are quantifier-eliminated recursively.
\item An existentially-quantified proposition (\cons{E} \bound{φ}); \bound{φ} is quantifier-eliminated recursively, and \field{elim} is applied to the result.
\item A universally-quantified proposition (\cons{A} \bound{φ}); \bound{φ} is quantifier-eliminated recursively, and the quantifier is treated as its (classical) existential dual (\func{\nope} \cons{E} \func{\nope}).\footnote{The validity of this under a constructive metatheory is not immediately obvious, and will be addressed Section~\ref{subsec:qecor}.}

\end{enumerate}
This procedure is formalized as the function \func{lift-qe}:
\begin{code}%
\>[0]\AgdaIndent{4}{}\<[4]%
\>[4]\AgdaFunction{lift-qe} \AgdaSymbol{:} \AgdaSymbol{\{}\AgdaBound{n} \AgdaSymbol{:} \AgdaDatatype{ℕ}\AgdaSymbol{\}} \AgdaSymbol{→} \AgdaRecord{QE} \AgdaSymbol{→} \AgdaDatatype{Prop} \AgdaBound{n} \AgdaSymbol{→} \AgdaDatatype{Prop} \AgdaBound{n}\<%
\\
\>[0]\AgdaIndent{4}{}\<[4]%
\>[4]\AgdaFunction{lift-qe-qfree} \AgdaSymbol{:} \AgdaSymbol{\{}\AgdaBound{n} \AgdaSymbol{:} \AgdaDatatype{ℕ}\AgdaSymbol{\}} \AgdaSymbol{(}\AgdaBound{qe} \AgdaSymbol{:} \AgdaRecord{QE}\AgdaSymbol{)} \AgdaSymbol{(}\AgdaBound{φ} \AgdaSymbol{:} \AgdaDatatype{Prop} \AgdaBound{n}\AgdaSymbol{)} \AgdaSymbol{→} \AgdaDatatype{QFree} \AgdaSymbol{(}\AgdaFunction{lift-qe} \AgdaBound{qe} \AgdaBound{φ}\AgdaSymbol{)}\<%
\\
\\
\>[0]\AgdaIndent{4}{}\<[4]%
\>[4]\AgdaFunction{lift-qe} \AgdaSymbol{\_} \AgdaInductiveConstructor{⊥⊥} \<[5]%
\>[5]\AgdaSymbol{=} \AgdaInductiveConstructor{⊥⊥}\<%
\\
\>[0]\AgdaIndent{4}{}\<[4]%
\>[4]\AgdaFunction{lift-qe} \AgdaSymbol{\_} \AgdaSymbol{(}\AgdaInductiveConstructor{atom} \AgdaBound{a}\AgdaSymbol{)} \<[5]%
\>[5]\AgdaSymbol{=} \AgdaInductiveConstructor{atom} \AgdaBound{a}\<%
\\
\>[0]\AgdaIndent{4}{}\<[4]%
\>[4]\AgdaFunction{lift-qe} \AgdaBound{qe} \AgdaSymbol{(}\AgdaBound{φ₁} \AgdaInductiveConstructor{∨} \AgdaBound{φ₂}\AgdaSymbol{)} \<[5]%
\>[5]\AgdaSymbol{=} \AgdaSymbol{(}\AgdaFunction{lift-qe} \AgdaBound{qe} \AgdaBound{φ₁}\AgdaSymbol{)} \AgdaInductiveConstructor{∨} \AgdaSymbol{(}\AgdaFunction{lift-qe} \AgdaBound{qe} \AgdaBound{φ₂}\AgdaSymbol{)}\<%
\\
\>[0]\AgdaIndent{4}{}\<[4]%
\>[4]\AgdaFunction{lift-qe} \AgdaBound{qe} \AgdaSymbol{(}\AgdaBound{φ₁} \AgdaInductiveConstructor{∧} \AgdaBound{φ₂}\AgdaSymbol{)} \<[5]%
\>[5]\AgdaSymbol{=} \AgdaSymbol{(}\AgdaFunction{lift-qe} \AgdaBound{qe} \AgdaBound{φ₁}\AgdaSymbol{)} \AgdaInductiveConstructor{∧} \AgdaSymbol{(}\AgdaFunction{lift-qe} \AgdaBound{qe} \AgdaBound{φ₂}\AgdaSymbol{)}\<%
\\
\>[0]\AgdaIndent{4}{}\<[4]%
\>[4]\AgdaFunction{lift-qe} \AgdaBound{qe} \AgdaSymbol{(}\AgdaBound{φ₁} \AgdaInductiveConstructor{⇒} \AgdaBound{φ₂}\AgdaSymbol{)} \<[5]%
\>[5]\AgdaSymbol{=} \AgdaSymbol{(}\AgdaFunction{lift-qe} \AgdaBound{qe} \AgdaBound{φ₁}\AgdaSymbol{)} \AgdaInductiveConstructor{⇒} \AgdaSymbol{(}\AgdaFunction{lift-qe} \AgdaBound{qe} \AgdaBound{φ₂}\AgdaSymbol{)}\<%
\\
\>[0]\AgdaIndent{4}{}\<[4]%
\>[4]\AgdaFunction{lift-qe} \AgdaBound{qe} \AgdaSymbol{(}\AgdaInductiveConstructor{E} \AgdaBound{φ}\AgdaSymbol{)} \<[5]%
\>[5]\AgdaSymbol{=} \AgdaField{QE.elim} \AgdaBound{qe} \AgdaSymbol{(}\AgdaFunction{lift-qe} \AgdaBound{qe} \AgdaBound{φ}\AgdaSymbol{)} \AgdaSymbol{(}\AgdaFunction{lift-qe-qfree} \AgdaBound{qe} \AgdaBound{φ}\AgdaSymbol{)}\<%
\\
\>[0]\AgdaIndent{4}{}\<[4]%
\>[4]\AgdaFunction{lift-qe} \AgdaBound{qe} \AgdaSymbol{(}\AgdaInductiveConstructor{A} \AgdaBound{φ}\AgdaSymbol{)} \<[5]%
\>[5]\AgdaSymbol{=} \AgdaFunction{\textasciitilde{}} \AgdaSymbol{(}\AgdaField{QE.elim} \AgdaBound{qe} \AgdaSymbol{(}\AgdaFunction{\textasciitilde{}} \AgdaFunction{lift-qe} \AgdaBound{qe} \AgdaBound{φ}\AgdaSymbol{)} \AgdaSymbol{(}\AgdaFunction{\textasciitilde{}-qf} \AgdaFunction{lift-qe-qfree} \AgdaBound{qe} \AgdaBound{φ}\AgdaSymbol{))}\<%
\end{code}

The function \func{lift-qe-qfree} (contents omitted) affirms that \func{lift-qe} does indeed eliminate quantifiers, via recursion on the proposition's structure and the use of \field{QE.qfree}.

\subsubsection{Correctness}\label{subsec:qecor}

The correctness of the lifted procedure---that \func{lift-qe} \bound{qe} \bound{φ} is equivalent to \bound{φ}---is proven recursively based on the correctess of the single-step procedure.
This takes the form of two functions, proving each direction of the equivalence:

\begin{code}%
\>[0]\AgdaIndent{4}{}\<[4]%
\>[4]\AgdaFunction{lift-qe-fwd} \<[5]%
\>[5]\AgdaSymbol{:} \AgdaSymbol{\{}\AgdaBound{n} \AgdaSymbol{:} \AgdaDatatype{ℕ}\AgdaSymbol{\}} \AgdaSymbol{(}\AgdaBound{qe} \AgdaSymbol{:} \AgdaRecord{QE}\AgdaSymbol{)} \AgdaSymbol{(}\AgdaBound{φ} \AgdaSymbol{:} \AgdaDatatype{Prop} \AgdaBound{n}\AgdaSymbol{)} \AgdaSymbol{(}\AgdaBound{e} \AgdaSymbol{:} \AgdaDatatype{Vec} \AgdaField{Y} \AgdaBound{n}\AgdaSymbol{)} \AgdaSymbol{→} \AgdaFunction{⟦} \AgdaBound{φ} \AgdaFunction{⟧} \AgdaBound{e} \AgdaSymbol{→} \AgdaFunction{⟦} \AgdaFunction{lift-qe} \AgdaBound{qe} \AgdaBound{φ} \AgdaFunction{⟧} \AgdaBound{e}\<%
\\
\>[0]\AgdaIndent{4}{}\<[4]%
\>[4]\AgdaFunction{lift-qe-bwd} \<[5]%
\>[5]\AgdaSymbol{:} \AgdaSymbol{\{}\AgdaBound{n} \AgdaSymbol{:} \AgdaDatatype{ℕ}\AgdaSymbol{\}} \AgdaSymbol{(}\AgdaBound{qe} \AgdaSymbol{:} \AgdaRecord{QE}\AgdaSymbol{)} \AgdaSymbol{(}\AgdaBound{φ} \AgdaSymbol{:} \AgdaDatatype{Prop} \AgdaBound{n}\AgdaSymbol{)} \AgdaSymbol{(}\AgdaBound{e} \AgdaSymbol{:} \AgdaDatatype{Vec} \AgdaField{Y} \AgdaBound{n}\AgdaSymbol{)} \AgdaSymbol{→} \AgdaFunction{⟦} \AgdaFunction{lift-qe} \AgdaBound{qe} \AgdaBound{φ} \AgdaFunction{⟧} \AgdaBound{e} \AgdaSymbol{→} \AgdaFunction{⟦} \AgdaBound{φ} \AgdaFunction{⟧} \AgdaBound{e}\<%
\end{code}

For both directions, the cases \cons{⊥⊥} and \cons{atom} are trivial; the former is impossible and the latter is unchanged by \func{lift-qe}.
For \cons{∨}, \cons{∧}, and \cons{⇒}, correctness of \func{lift-qe} is proven recursively on each sub-proposition, and then combined:

\begin{code}%
\>[0]\AgdaIndent{4}{}\<[4]%
\>[4]\AgdaFunction{lift-qe-fwd} \AgdaBound{qe} \AgdaSymbol{(}\AgdaBound{φ₁} \AgdaInductiveConstructor{∨} \AgdaBound{φ₂}\AgdaSymbol{)} \AgdaBound{e} \<[10]%
\>[10]\AgdaSymbol{=} \AgdaFunction{Sum.map} \AgdaSymbol{(}\AgdaFunction{lift-qe-fwd} \AgdaBound{qe} \AgdaBound{φ₁} \AgdaBound{e}\AgdaSymbol{)} \AgdaSymbol{(}\AgdaFunction{lift-qe-fwd} \AgdaBound{qe} \AgdaBound{φ₂} \AgdaBound{e}\AgdaSymbol{)}\<%
\\
\>[0]\AgdaIndent{4}{}\<[4]%
\>[4]\AgdaFunction{lift-qe-fwd} \AgdaBound{qe} \AgdaSymbol{(}\AgdaBound{φ₁} \AgdaInductiveConstructor{∧} \AgdaBound{φ₂}\AgdaSymbol{)} \AgdaBound{e} \<[10]%
\>[10]\AgdaSymbol{=} \AgdaFunction{Product.map} \AgdaSymbol{(}\AgdaFunction{lift-qe-fwd} \AgdaBound{qe} \AgdaBound{φ₁} \AgdaBound{e}\AgdaSymbol{)} \AgdaSymbol{(}\AgdaFunction{lift-qe-fwd} \AgdaBound{qe} \AgdaBound{φ₂} \AgdaBound{e}\AgdaSymbol{)}\<%
\\
\>[0]\AgdaIndent{4}{}\<[4]%
\>[4]\AgdaFunction{lift-qe-fwd} \AgdaBound{qe} \AgdaSymbol{(}\AgdaBound{φ₁} \AgdaInductiveConstructor{⇒} \AgdaBound{φ₂}\AgdaSymbol{)} \AgdaBound{e} \<[10]%
\>[10]\AgdaSymbol{=} \AgdaSymbol{λ} \AgdaBound{f} \AgdaSymbol{→} \AgdaFunction{lift-qe-fwd} \AgdaBound{qe} \AgdaBound{φ₂} \AgdaBound{e} \AgdaFunction{∘} \AgdaBound{f} \AgdaFunction{∘} \AgdaFunction{lift-qe-bwd} \AgdaBound{qe} \AgdaBound{φ₁} \AgdaBound{e}\<%
\\
\\
\>[0]\AgdaIndent{4}{}\<[4]%
\>[4]\AgdaFunction{lift-qe-bwd} \AgdaBound{qe} \AgdaSymbol{(}\AgdaBound{φ₁} \AgdaInductiveConstructor{∨} \AgdaBound{φ₂}\AgdaSymbol{)} \AgdaBound{e} \<[10]%
\>[10]\AgdaSymbol{=} \AgdaFunction{Sum.map} \AgdaSymbol{(}\AgdaFunction{lift-qe-bwd} \AgdaBound{qe} \AgdaBound{φ₁} \AgdaBound{e}\AgdaSymbol{)} \AgdaSymbol{(}\AgdaFunction{lift-qe-bwd} \AgdaBound{qe} \AgdaBound{φ₂} \AgdaBound{e}\AgdaSymbol{)}\<%
\\
\>[0]\AgdaIndent{4}{}\<[4]%
\>[4]\AgdaFunction{lift-qe-bwd} \AgdaBound{qe} \AgdaSymbol{(}\AgdaBound{φ₁} \AgdaInductiveConstructor{∧} \AgdaBound{φ₂}\AgdaSymbol{)} \AgdaBound{e} \<[10]%
\>[10]\AgdaSymbol{=} \AgdaFunction{Product.map} \AgdaSymbol{(}\AgdaFunction{lift-qe-bwd} \AgdaBound{qe} \AgdaBound{φ₁} \AgdaBound{e}\AgdaSymbol{)} \AgdaSymbol{(}\AgdaFunction{lift-qe-bwd} \AgdaBound{qe} \AgdaBound{φ₂} \AgdaBound{e}\AgdaSymbol{)}\<%
\\
\>[0]\AgdaIndent{4}{}\<[4]%
\>[4]\AgdaFunction{lift-qe-bwd} \AgdaBound{qe} \AgdaSymbol{(}\AgdaBound{φ₁} \AgdaInductiveConstructor{⇒} \AgdaBound{φ₂}\AgdaSymbol{)} \AgdaBound{e} \<[10]%
\>[10]\AgdaSymbol{=} \AgdaSymbol{λ} \AgdaBound{f} \AgdaSymbol{→} \AgdaFunction{lift-qe-bwd} \AgdaBound{qe} \AgdaBound{φ₂} \AgdaBound{e} \AgdaFunction{∘} \AgdaBound{f} \AgdaFunction{∘} \AgdaFunction{lift-qe-fwd} \AgdaBound{qe} \AgdaBound{φ₁} \AgdaBound{e}\<%
\end{code}

In the case of existential quantification, \func{lift-qe} recurses on \bound{φ}, producing an equivalent, quantifier-free \bound{ψ}, which \record{QE.elim} is applied to.
The reasoning behind this is as follows:

$$\exists x. \phi \iff \exists x.\psi \iff elim(\psi).$$

The first equivalence is justified by the correctness of \func{lift-qe} on \bound{φ}, obtained recursively, and the second by the correctness of the the single-step procedure, given by \field{QE.equiv}.
Formalized:

\begin{code}%
\>[0]\AgdaIndent{4}{}\<[4]%
\>[4]\AgdaFunction{lift-qe-fwd} \AgdaBound{qe} \AgdaSymbol{(}\AgdaInductiveConstructor{E} \AgdaBound{φ}\AgdaSymbol{)} \AgdaBound{e} \<[27]%
\>[27]\<%
\\
\>[4]\AgdaIndent{6}{}\<[6]%
\>[6]\AgdaSymbol{=} \AgdaField{proj₁} \AgdaSymbol{(}\AgdaField{QE.equiv} \AgdaBound{qe} \AgdaSymbol{(}\AgdaFunction{lift-qe} \AgdaBound{qe} \AgdaBound{φ}\AgdaSymbol{)} \AgdaSymbol{(}\AgdaFunction{lift-qe-qfree} \AgdaBound{qe} \AgdaBound{φ}\AgdaSymbol{)} \AgdaBound{e}\AgdaSymbol{)}\<%
\\
\>[6]\AgdaIndent{8}{}\<[8]%
\>[8]\AgdaFunction{∘} \AgdaFunction{Σ-map} \AgdaSymbol{(λ} \AgdaBound{y} \AgdaSymbol{→} \AgdaFunction{lift-qe-fwd} \AgdaBound{qe} \AgdaBound{φ} \AgdaSymbol{(}\AgdaBound{y} \AgdaInductiveConstructor{∷} \AgdaBound{e}\AgdaSymbol{))}\<%
\\
\\
\>[0]\AgdaIndent{4}{}\<[4]%
\>[4]\AgdaFunction{lift-qe-bwd} \AgdaBound{qe} \AgdaSymbol{(}\AgdaInductiveConstructor{E} \AgdaBound{φ}\AgdaSymbol{)} \AgdaBound{e}\<%
\\
\>[4]\AgdaIndent{6}{}\<[6]%
\>[6]\AgdaSymbol{=} \AgdaFunction{Σ-map} \AgdaSymbol{(λ} \AgdaBound{y} \AgdaSymbol{→} \AgdaFunction{lift-qe-bwd} \AgdaBound{qe} \AgdaBound{φ} \AgdaSymbol{(}\AgdaBound{y} \AgdaInductiveConstructor{∷} \AgdaBound{e}\AgdaSymbol{))}\<%
\\
\>[6]\AgdaIndent{8}{}\<[8]%
\>[8]\AgdaFunction{∘} \AgdaField{proj₂} \AgdaSymbol{(}\AgdaField{QE.equiv} \AgdaBound{qe} \AgdaSymbol{(}\AgdaFunction{lift-qe} \AgdaBound{qe} \AgdaBound{φ}\AgdaSymbol{)} \AgdaSymbol{(}\AgdaFunction{lift-qe-qfree} \AgdaBound{qe} \AgdaBound{φ}\AgdaSymbol{)} \AgdaBound{e}\AgdaSymbol{)}\<%
\end{code}
where \func{Σ-map} proves that if $\forall x.(B(x) \impl C(x))$, then $\exists x.B(x) \impl \exists x.C(x)$, in this case used to obtain $\exists x.\phi \iff \exists x.\psi$ from $\phi \iff \psi$:
\begin{code}%
\>[0]\AgdaIndent{4}{}\<[4]%
\>[4]\AgdaFunction{Σ-map} \AgdaSymbol{:} \AgdaSymbol{\{}\AgdaBound{A} \AgdaSymbol{:} \AgdaPrimitiveType{Set}\AgdaSymbol{\}} \AgdaSymbol{\{}\AgdaBound{B} \AgdaBound{C} \AgdaSymbol{:} \AgdaBound{A} \AgdaSymbol{→} \AgdaPrimitiveType{Set}\AgdaSymbol{\}} \AgdaSymbol{→} \AgdaSymbol{((}\AgdaBound{a} \AgdaSymbol{:} \AgdaBound{A}\AgdaSymbol{)} \AgdaSymbol{→} \AgdaBound{B} \AgdaBound{a} \AgdaSymbol{→} \AgdaBound{C} \AgdaBound{a}\AgdaSymbol{)} \AgdaSymbol{→} \AgdaRecord{Σ} \AgdaBound{A} \AgdaBound{B} \AgdaSymbol{→} \AgdaRecord{Σ} \AgdaBound{A} \AgdaBound{C}\<%
\\
\>[0]\AgdaIndent{4}{}\<[4]%
\>[4]\AgdaFunction{Σ-map} \AgdaBound{f} \AgdaSymbol{(}\AgdaBound{a} \AgdaInductiveConstructor{,} \AgdaBound{b}\AgdaSymbol{)} \AgdaSymbol{=} \AgdaSymbol{(}\AgdaBound{a} \AgdaInductiveConstructor{,} \AgdaBound{f} \AgdaBound{a} \AgdaBound{b}\AgdaSymbol{)}\<%
\end{code}

The case of universal quantification is cause for mild concern, however: \func{lift-qe} treats the quantifier \cons{A} as its classical dual \func{\nope} \cons{E} \func{\nope}.

In a classical metatheory, correctness could be obtained as follows (once again taking $\psi$ to be the quantifier-free equivalent of $\phi$):
$$\forall x. \phi \iff \neg \exists x. \neg \phi \iff \neg \exists x. \neg \psi \iff \neg elim(\neg \psi).$$
The first equivalence is justified by quantifier duality, the second by the correctness of \func{lift-qe} on \bound{φ} ($\phi \iff \psi$, obtainable via recursion), and the third by the correctness of \field{QE.elim} (\field{QE.equiv}).
Conceptually, this corresponds to treating $\forall$ as $\neg\exists\neg$ from the outset.

Under a constructive metatheory, though, the first equivalence is not valid due to the lack of complete quantifier duality: while $\forall x.\phi \impl \neg \exists x.\neg \phi$, the converse is not provable.
However, in this case this can be neatly sidestepped by rearranging things slightly:
$$\forall x. \phi \iff \forall x. \psi \iff \neg \exists x. \neg \psi \iff \neg elim(\neg \psi).$$
The difference here is that the quantifier duality is applied to $\psi$, instead of $\phi$.
$\psi$, being quantifier-free, has decidable semantics (by \func{qfree-dec}), and as a consequence the necessary quantifier duality can in fact be proven.
General forms of the duality are formalized as follows:
\begin{code}%
\>[0]\AgdaIndent{4}{}\<[4]%
\>[4]\AgdaFunction{∀-duality-fwd} \AgdaSymbol{:} \AgdaSymbol{\{}\AgdaBound{A} \AgdaSymbol{:} \AgdaPrimitiveType{Set}\AgdaSymbol{\}} \AgdaSymbol{\{}\AgdaBound{B} \AgdaSymbol{:} \AgdaBound{A} \AgdaSymbol{→} \AgdaPrimitiveType{Set}\AgdaSymbol{\}} \AgdaSymbol{→} \AgdaSymbol{((}\AgdaBound{a} \AgdaSymbol{:} \AgdaBound{A}\AgdaSymbol{)} \AgdaSymbol{→} \AgdaBound{B} \AgdaBound{a}\AgdaSymbol{)} \AgdaSymbol{→} \AgdaFunction{¬} \AgdaRecord{Σ} \AgdaBound{A} \AgdaSymbol{(}\AgdaFunction{¬\_} \AgdaFunction{∘} \AgdaBound{B}\AgdaSymbol{)}\<%
\\
\>[0]\AgdaIndent{4}{}\<[4]%
\>[4]\AgdaFunction{∀-duality-fwd} \AgdaBound{all-true} \AgdaSymbol{(}\AgdaBound{a} \AgdaInductiveConstructor{,} \AgdaBound{is-false}\AgdaSymbol{)} \AgdaSymbol{=} \AgdaBound{is-false} \AgdaSymbol{(}\AgdaBound{all-true} \AgdaBound{a}\AgdaSymbol{)}\<%
\\
\\
\>[0]\AgdaIndent{4}{}\<[4]%
\>[4]\AgdaFunction{∀-duality-bwd} \AgdaSymbol{:} \AgdaSymbol{\{}\AgdaBound{A} \AgdaSymbol{:} \AgdaPrimitiveType{Set}\AgdaSymbol{\}} \AgdaSymbol{\{}\AgdaBound{B} \AgdaSymbol{:} \AgdaBound{A} \AgdaSymbol{→} \AgdaPrimitiveType{Set}\AgdaSymbol{\}} \AgdaSymbol{→} \AgdaSymbol{((}\AgdaBound{a} \AgdaSymbol{:} \AgdaBound{A}\AgdaSymbol{)} \AgdaSymbol{→} \AgdaDatatype{Dec} \AgdaSymbol{(}\AgdaBound{B} \AgdaBound{a}\AgdaSymbol{))} \AgdaSymbol{→}\<%
\\
\>[4]\AgdaIndent{6}{}\<[6]%
\>[6]\AgdaFunction{¬} \AgdaRecord{Σ} \AgdaBound{A} \AgdaSymbol{(}\AgdaFunction{¬\_} \AgdaFunction{∘} \AgdaBound{B}\AgdaSymbol{)} \AgdaSymbol{→} \AgdaSymbol{((}\AgdaBound{a} \AgdaSymbol{:} \AgdaBound{A}\AgdaSymbol{)} \AgdaSymbol{→} \AgdaBound{B} \AgdaBound{a}\AgdaSymbol{)}\<%
\\
\>[0]\AgdaIndent{4}{}\<[4]%
\>[4]\AgdaFunction{∀-duality-bwd} \AgdaBound{decide} \AgdaBound{none-false} \AgdaBound{a} \AgdaKeyword{with} \AgdaBound{decide} \AgdaBound{a}\<%
\\
\>[0]\AgdaIndent{4}{}\<[4]%
\>[4]\AgdaSymbol{...} \AgdaSymbol{|} \AgdaInductiveConstructor{yes} \AgdaBound{a-true} \AgdaSymbol{=} \AgdaBound{a-true}\<%
\\
\>[0]\AgdaIndent{4}{}\<[4]%
\>[4]\AgdaSymbol{...} \AgdaSymbol{|} \AgdaInductiveConstructor{no} \AgdaBound{a-false} \AgdaSymbol{=} \AgdaFunction{⊥-elim} \AgdaSymbol{(}\AgdaBound{none-false} \AgdaSymbol{(}\AgdaBound{a} \AgdaInductiveConstructor{,} \AgdaBound{a-false}\AgdaSymbol{))}\<%
\end{code}
It is noted that the ``backward'' direction requires that \bound{B} be decidable.\footnote{While it could have been formulated to use the weaker requirement that \func{¬} \func{¬} \bound{B} \bound{a} \asym{→} \bound{B} \bound{a}, there is no particular benefit to doing so in this case.}
The correctness proof then proceeds as outlined above:
\begin{code}%
\>[0]\AgdaIndent{4}{}\<[4]%
\>[4]\AgdaFunction{lift-qe-fwd} \AgdaBound{qe} \AgdaSymbol{(}\AgdaInductiveConstructor{A} \AgdaBound{φ}\AgdaSymbol{)} \AgdaBound{e}\<%
\\
\>[4]\AgdaIndent{6}{}\<[6]%
\>[6]\AgdaSymbol{=} \AgdaFunction{contraposition} \AgdaSymbol{(}\AgdaField{proj₂} \AgdaSymbol{(}\AgdaField{QE.equiv} \AgdaBound{qe} \AgdaSymbol{(}\AgdaFunction{\textasciitilde{}} \AgdaFunction{lift-qe} \AgdaBound{qe} \AgdaBound{φ}\AgdaSymbol{)} \AgdaSymbol{(}\AgdaFunction{\textasciitilde{}-qf} \AgdaFunction{lift-qe-qfree} \AgdaBound{qe} \AgdaBound{φ}\AgdaSymbol{)} \AgdaBound{e}\AgdaSymbol{))}\<%
\\
\>[0]\AgdaIndent{8}{}\<[8]%
\>[8]\AgdaFunction{∘} \AgdaFunction{∀-duality-fwd}\<%
\\
\>[0]\AgdaIndent{8}{}\<[8]%
\>[8]\AgdaFunction{∘} \AgdaFunction{Π-map} \AgdaSymbol{(λ} \AgdaBound{y} \AgdaSymbol{→} \AgdaFunction{lift-qe-fwd} \AgdaBound{qe} \AgdaBound{φ} \AgdaSymbol{(}\AgdaBound{y} \AgdaInductiveConstructor{∷} \AgdaBound{e}\AgdaSymbol{))}\<%
\\
\\
\>[0]\AgdaIndent{4}{}\<[4]%
\>[4]\AgdaFunction{lift-qe-bwd} \AgdaBound{qe} \AgdaSymbol{(}\AgdaInductiveConstructor{A} \AgdaBound{φ}\AgdaSymbol{)} \AgdaBound{e}\<%
\\
\>[4]\AgdaIndent{6}{}\<[6]%
\>[6]\AgdaSymbol{=} \AgdaFunction{Π-map} \AgdaSymbol{(λ} \AgdaBound{y} \AgdaSymbol{→} \AgdaFunction{lift-qe-bwd} \AgdaBound{qe} \AgdaBound{φ} \AgdaSymbol{(}\AgdaBound{y} \AgdaInductiveConstructor{∷} \AgdaBound{e}\AgdaSymbol{))}\<%
\\
\>[6]\AgdaIndent{8}{}\<[8]%
\>[8]\AgdaFunction{∘} \AgdaFunction{∀-duality-bwd} \AgdaSymbol{(λ} \AgdaBound{y} \AgdaSymbol{→} \AgdaFunction{qfree-dec} \AgdaSymbol{(}\AgdaFunction{lift-qe} \AgdaBound{qe} \AgdaBound{φ}\AgdaSymbol{)} \AgdaSymbol{(}\AgdaFunction{lift-qe-qfree} \AgdaBound{qe} \AgdaBound{φ}\AgdaSymbol{)} \AgdaSymbol{(}\AgdaBound{y} \AgdaInductiveConstructor{∷} \AgdaBound{e}\AgdaSymbol{))}\<%
\\
\>[0]\AgdaIndent{8}{}\<[8]%
\>[8]\AgdaFunction{∘} \AgdaFunction{contraposition} \AgdaSymbol{(}\AgdaField{proj₁} \AgdaSymbol{(}\AgdaField{QE.equiv} \AgdaBound{qe} \AgdaSymbol{(}\AgdaFunction{\textasciitilde{}} \AgdaFunction{lift-qe} \AgdaBound{qe} \AgdaBound{φ}\AgdaSymbol{)} \AgdaSymbol{(}\AgdaFunction{\textasciitilde{}-qf} \AgdaFunction{lift-qe-qfree} \AgdaBound{qe} \AgdaBound{φ}\AgdaSymbol{)} \AgdaBound{e}\AgdaSymbol{))}\<%
\end{code}
where \func{Π-map} is the dependent product/universal quantification counterpart of \func{Σ-map}:
\begin{code}%
\>[0]\AgdaIndent{4}{}\<[4]%
\>[4]\AgdaFunction{Π-map} \AgdaSymbol{:} \AgdaSymbol{\{}\AgdaBound{A} \AgdaSymbol{:} \AgdaPrimitiveType{Set}\AgdaSymbol{\}} \AgdaSymbol{\{}\AgdaBound{B} \AgdaBound{C} \AgdaSymbol{:} \AgdaBound{A} \AgdaSymbol{→} \AgdaPrimitiveType{Set}\AgdaSymbol{\}} \AgdaSymbol{→}\<%
\\
\>[4]\AgdaIndent{6}{}\<[6]%
\>[6]\AgdaSymbol{((}\AgdaBound{a} \AgdaSymbol{:} \AgdaBound{A}\AgdaSymbol{)} \AgdaSymbol{→} \AgdaBound{B} \AgdaBound{a} \AgdaSymbol{→} \AgdaBound{C} \AgdaBound{a}\AgdaSymbol{)} \AgdaSymbol{→} \AgdaSymbol{((}\AgdaBound{a} \AgdaSymbol{:} \AgdaBound{A}\AgdaSymbol{)} \AgdaSymbol{→} \AgdaBound{B} \AgdaBound{a}\AgdaSymbol{)} \AgdaSymbol{→} \AgdaSymbol{((}\AgdaBound{a} \AgdaSymbol{:} \AgdaBound{A}\AgdaSymbol{)} \AgdaSymbol{→} \AgdaBound{C} \AgdaBound{a}\AgdaSymbol{)}\<%
\\
\>[0]\AgdaIndent{4}{}\<[4]%
\>[4]\AgdaFunction{Π-map} \AgdaBound{f} \AgdaBound{g} \AgdaBound{a} \AgdaSymbol{=} \AgdaBound{f} \AgdaBound{a} \AgdaSymbol{(}\AgdaBound{g} \AgdaBound{a}\AgdaSymbol{)}\<%
\end{code}

\subsection{Decidability}\label{subsec:dec}

Given a single-step elimination procedure \bound{qe} \asym{:} \record{QE}, the decidability of any proposition \bound{φ} follows: \func{lift-qe} \bound{qe} \bound{φ} produces an equivalent, quantifier-free proposition \bound{ψ}.
As such, \bound{ψ} is decidable (\func{qfree-dec}).
Because \bound{φ} and \bound{ψ} are semantically equivalent (\func{lift-qe-fwd}, \func{lift-qe-bwd}), this immediately results in the decidability of \bound{φ}.

\begin{code}%
\>[4]\AgdaIndent{6}{}\<[6]%
\>[6]\AgdaFunction{⟦\_⟧?} \AgdaSymbol{:} \AgdaSymbol{\{}\AgdaBound{n} \AgdaSymbol{:} \AgdaDatatype{ℕ}\AgdaSymbol{\}} \AgdaSymbol{→} \AgdaSymbol{(}\AgdaBound{φ} \AgdaSymbol{:} \AgdaDatatype{Prop} \AgdaBound{n}\AgdaSymbol{)} \AgdaSymbol{→} \AgdaSymbol{(}\AgdaBound{e} \AgdaSymbol{:} \AgdaDatatype{Vec} \AgdaField{Y} \AgdaBound{n}\AgdaSymbol{)} \AgdaSymbol{→} \AgdaDatatype{Dec} \AgdaSymbol{(}\AgdaFunction{⟦} \AgdaBound{φ} \AgdaFunction{⟧} \AgdaBound{e}\AgdaSymbol{)}\<%
\\
\>[4]\AgdaIndent{6}{}\<[6]%
\>[6]\AgdaFunction{⟦} \AgdaBound{φ} \AgdaFunction{⟧?} \AgdaBound{e} \AgdaKeyword{with} \AgdaFunction{qfree-dec} \AgdaSymbol{(}\AgdaFunction{lift-qe} \AgdaBound{qe} \AgdaBound{φ}\AgdaSymbol{)} \AgdaSymbol{(}\AgdaFunction{lift-qe-qfree} \AgdaBound{qe} \AgdaBound{φ}\AgdaSymbol{)} \AgdaBound{e}\<%
\\
\>[4]\AgdaIndent{6}{}\<[6]%
\>[6]\AgdaSymbol{...} \AgdaSymbol{|} \AgdaInductiveConstructor{yes} \AgdaBound{⟦ψ⟧} \AgdaSymbol{=} \AgdaInductiveConstructor{yes} \AgdaSymbol{(}\AgdaFunction{lift-qe-bwd} \AgdaBound{qe} \AgdaBound{φ} \AgdaBound{e} \AgdaBound{⟦ψ⟧}\AgdaSymbol{)}\<%
\\
\>[4]\AgdaIndent{6}{}\<[6]%
\>[6]\AgdaSymbol{...} \AgdaSymbol{|} \AgdaInductiveConstructor{no} \AgdaBound{¬⟦ψ⟧} \AgdaSymbol{=} \AgdaInductiveConstructor{no} \AgdaSymbol{(}\AgdaBound{¬⟦ψ⟧} \AgdaFunction{∘} \AgdaFunction{lift-qe-fwd} \AgdaBound{qe} \AgdaBound{φ} \AgdaBound{e}\AgdaSymbol{)}\<%
\end{code}
The theory in question, whatever it may be, is thus proven decidable.

\subsubsection{Consequences}
With decidability, the law of excluded middle can be proven for the semantics of \data{Prop}:

\begin{code}%
\>[4]\AgdaIndent{6}{}\<[6]%
\>[6]\AgdaFunction{LEM} \AgdaSymbol{:} \AgdaSymbol{\{}\AgdaBound{n} \AgdaSymbol{:} \AgdaDatatype{ℕ}\AgdaSymbol{\}} \AgdaSymbol{(}\AgdaBound{φ} \AgdaSymbol{:} \AgdaDatatype{Prop} \AgdaBound{n}\AgdaSymbol{)} \AgdaSymbol{(}\AgdaBound{e} \AgdaSymbol{:} \AgdaDatatype{Vec} \AgdaField{Y} \AgdaBound{n}\AgdaSymbol{)} \AgdaSymbol{→} \AgdaFunction{⟦} \AgdaBound{φ} \AgdaInductiveConstructor{∨} \AgdaSymbol{(}\AgdaFunction{\textasciitilde{}} \AgdaBound{φ}\AgdaSymbol{)} \AgdaFunction{⟧} \AgdaBound{e}\<%
\\
\>[4]\AgdaIndent{6}{}\<[6]%
\>[6]\AgdaFunction{LEM} \AgdaBound{φ} \AgdaBound{e} \AgdaKeyword{with} \AgdaFunction{⟦} \AgdaBound{φ} \AgdaFunction{⟧?} \AgdaBound{e}\<%
\\
\>[4]\AgdaIndent{6}{}\<[6]%
\>[6]\AgdaSymbol{...} \AgdaSymbol{|} \AgdaInductiveConstructor{yes} \AgdaBound{⟦φ⟧} \<[7]%
\>[7]\AgdaSymbol{=} \AgdaInductiveConstructor{inj₁} \AgdaBound{⟦φ⟧}\<%
\\
\>[4]\AgdaIndent{6}{}\<[6]%
\>[6]\AgdaSymbol{...} \AgdaSymbol{|} \AgdaInductiveConstructor{no} \AgdaBound{¬⟦φ⟧} \<[7]%
\>[7]\AgdaSymbol{=} \AgdaInductiveConstructor{inj₂} \AgdaBound{¬⟦φ⟧}\<%
\end{code}

Using the law of excluded middle, the hitherto unavailable classical results become provable---with all of the benefits of a constructive metatheory (see Section~\ref{subsec:constructive}).
For example, for any proposition \(\phi\), \((\forall x.\phi) \vee (\exists x.\neg \phi)\) is provable:
\begin{code}%
\>[4]\AgdaIndent{6}{}\<[6]%
\>[6]\AgdaFunction{∀-or-∃¬} \AgdaSymbol{:} \AgdaSymbol{\{}\AgdaBound{n} \AgdaSymbol{:} \AgdaDatatype{ℕ}\AgdaSymbol{\}} \AgdaSymbol{(}\AgdaBound{φ} \AgdaSymbol{:} \AgdaDatatype{Prop} \AgdaSymbol{(}\AgdaInductiveConstructor{suc} \AgdaBound{n}\AgdaSymbol{))} \AgdaSymbol{(}\AgdaBound{e} \AgdaSymbol{:} \AgdaDatatype{Vec} \AgdaField{Y} \AgdaBound{n}\AgdaSymbol{)} \AgdaSymbol{→} \AgdaFunction{⟦} \AgdaSymbol{(}\AgdaInductiveConstructor{A} \AgdaBound{φ}\AgdaSymbol{)} \AgdaInductiveConstructor{∨} \AgdaSymbol{(}\AgdaInductiveConstructor{E} \AgdaFunction{\textasciitilde{}} \AgdaBound{φ}\AgdaSymbol{)} \AgdaFunction{⟧} \AgdaBound{e}\<%
\\
\>[4]\AgdaIndent{6}{}\<[6]%
\>[6]\AgdaFunction{∀-or-∃¬} \AgdaBound{φ} \AgdaBound{e} \AgdaKeyword{with} \AgdaFunction{⟦} \AgdaInductiveConstructor{E} \AgdaSymbol{(}\AgdaFunction{\textasciitilde{}} \AgdaBound{φ}\AgdaSymbol{)} \AgdaFunction{⟧?} \AgdaBound{e}\<%
\\
\>[4]\AgdaIndent{6}{}\<[6]%
\>[6]\AgdaSymbol{...} \AgdaSymbol{|} \AgdaInductiveConstructor{yes} \AgdaBound{⟦E\textasciitilde{}φ⟧} \AgdaSymbol{=} \AgdaInductiveConstructor{inj₂} \AgdaBound{⟦E\textasciitilde{}φ⟧}\<%
\\
\>[4]\AgdaIndent{6}{}\<[6]%
\>[6]\AgdaSymbol{...} \AgdaSymbol{|} \AgdaInductiveConstructor{no} \AgdaBound{¬⟦E\textasciitilde{}φ⟧}\<%
\\
\>[6]\AgdaIndent{8}{}\<[8]%
\>[8]\AgdaSymbol{=} \AgdaInductiveConstructor{inj₁} \AgdaSymbol{(λ} \AgdaBound{y} \AgdaSymbol{→} \AgdaFunction{[} \AgdaFunction{id} \AgdaFunction{,} \AgdaSymbol{(λ} \AgdaBound{¬⟦φ⟧} \AgdaSymbol{→} \AgdaFunction{contradiction} \AgdaSymbol{(}\AgdaBound{y} \AgdaInductiveConstructor{,} \AgdaBound{¬⟦φ⟧}\AgdaSymbol{)} \AgdaBound{¬⟦E\textasciitilde{}φ⟧}\AgdaSymbol{)} \AgdaFunction{]′} \AgdaSymbol{(}\AgdaFunction{LEM} \AgdaBound{φ} \AgdaSymbol{(}\AgdaBound{y} \AgdaInductiveConstructor{∷} \AgdaBound{e}\AgdaSymbol{)))}\<%
\end{code}
While trivial in a classical setting, in a constructive setting this produces either \emph{(i)} a proof that \(\phi\) is true for every \(x\), or \emph{(ii)} a counterexample---a value for \(x\) which causes \(\phi\) to be false (and a proof thereof).
Analogous results are obtained for the theorem \((\exists x.\phi) \vee (\forall x. \neg \phi)\).

\subsection{Disjunctive Normal Form and Products}\label{subsec:dnf}

The most basic formulation of a single-step elimination procedure (\record{QE}) is one that accepts a quantifier-free proposition $\phi$, and produces a quantifier-free proposition $\psi$ such that $(\exists x.\phi) \iff \psi$.
There are no restrictions on the form of $\phi$, other than that it is quantifier-free.
In practice, many quantifier elimination procedures require that $\phi$ be transformed into a special form first, as seen in the works of Herbrand~\cite{herbrand}, Nipkow~\cite{nipkowjar}, and Allais~\cite{allaisgit}.

As discussed in Section~\ref{subsec:bkgqe} one such form is Disjunctive Normal Form (DNF), where propositions take the shape of a disjunction of conjunctions of literals.
The utility of this form is that since existential quantification distributes across disjunction, the problem of quantifier elimination on DNF formulae trivially reduces to quantifier elimination on conjunctions of literals:
\[\exists x.\phi \iff \exists x. (C_1 \vee C_2 \vee \ldots \vee C_n) \iff (\exists x. C_1) \vee (\exists x. C_2) \vee \ldots \vee (\exists x. C_n).\]

The mechanics and correctness of the conversion to DNF are not discussed here, as they do not differ significantly from under a classical metatheory.\footnote{Recall that the semantics of an atom is decidable, so De Morgan's laws hold.} The results are a function \func{dnf} that transforms any proposition into DNF, and proof of its correctness:
\begin{code}%
\>[0]\AgdaIndent{4}{}\<[4]%
\>[4]\AgdaFunction{dnf} \<[5]%
\>[5]\AgdaSymbol{:} \AgdaSymbol{\{}\AgdaBound{n} \AgdaSymbol{:} \AgdaDatatype{ℕ}\AgdaSymbol{\}} \AgdaSymbol{(}\AgdaBound{p} \AgdaSymbol{:} \AgdaDatatype{Prop} \AgdaBound{n}\AgdaSymbol{)} \AgdaSymbol{(}\AgdaBound{qf} \AgdaSymbol{:} \AgdaDatatype{QFree} \AgdaBound{p}\AgdaSymbol{)} \AgdaSymbol{→} \AgdaFunction{DNF} \AgdaBound{n}\<%
\\
\>[4]\AgdaFunction{dnf-fwd} \<[5]%
\>[5]\AgdaSymbol{:} \AgdaSymbol{\{}\AgdaBound{n} \AgdaSymbol{:} \AgdaDatatype{ℕ}\AgdaSymbol{\}} \AgdaSymbol{(}\AgdaBound{p} \AgdaSymbol{:} \AgdaDatatype{Prop} \AgdaBound{n}\AgdaSymbol{)} \AgdaSymbol{(}\AgdaBound{qf} \AgdaSymbol{:} \AgdaDatatype{QFree} \AgdaBound{p}\AgdaSymbol{)} \AgdaSymbol{(}\AgdaBound{e} \AgdaSymbol{:} \AgdaDatatype{Vec} \AgdaField{Y} \AgdaBound{n}\AgdaSymbol{)} \AgdaSymbol{→} \AgdaFunction{⟦} \AgdaBound{p} \AgdaFunction{⟧} \AgdaBound{e} \AgdaSymbol{→} \AgdaFunction{⟦} \AgdaFunction{D.i} \AgdaSymbol{(}\AgdaFunction{dnf} \AgdaBound{p} \AgdaBound{qf}\AgdaSymbol{)} \AgdaFunction{⟧} \AgdaBound{e}\<%
\\
\>[0]\AgdaIndent{4}{}\<[4]%
\>[4]\AgdaFunction{dnf-bwd} \<[5]%
\>[5]\AgdaSymbol{:} \AgdaSymbol{\{}\AgdaBound{n} \AgdaSymbol{:} \AgdaDatatype{ℕ}\AgdaSymbol{\}} \AgdaSymbol{(}\AgdaBound{p} \AgdaSymbol{:} \AgdaDatatype{Prop} \AgdaBound{n}\AgdaSymbol{)} \AgdaSymbol{(}\AgdaBound{qf} \AgdaSymbol{:} \AgdaDatatype{QFree} \AgdaBound{p}\AgdaSymbol{)} \AgdaSymbol{(}\AgdaBound{e} \AgdaSymbol{:} \AgdaDatatype{Vec} \AgdaField{Y} \AgdaBound{n}\AgdaSymbol{)} \AgdaSymbol{→} \AgdaFunction{⟦} \AgdaFunction{D.i} \AgdaSymbol{(}\AgdaFunction{dnf} \AgdaBound{p} \AgdaBound{qf}\AgdaSymbol{)} \AgdaFunction{⟧} \AgdaBound{e} \AgdaSymbol{→} \AgdaFunction{⟦} \AgdaBound{p} \AgdaFunction{⟧} \AgdaBound{e}\<%
\end{code}

The \func{DNF} datatype employed above is a list of lists of literals, which can be ``interpreted'' as a proposition using the function \func{D.i}.\footnote{An alternative approach is to use an actual \data{Prop}, along with a proof that it is in disjunctive normal form, but this can make manipulation cumbersome.}

Single-step elimination on propositions in DNF is defined much like the more general \record{QE}:

\begin{code}%
\>[0]\AgdaIndent{4}{}\<[4]%
\>[4]\AgdaKeyword{record} \AgdaRecord{DNFQE} \AgdaSymbol{:} \AgdaPrimitiveType{Set} \AgdaKeyword{where}\<%
\\
\>[4]\AgdaIndent{6}{}\<[6]%
\>[6]\AgdaKeyword{field}\<%
\\
\>[6]\AgdaIndent{8}{}\<[8]%
\>[8]\AgdaField{elim} \<[14]%
\>[14]\AgdaSymbol{:} \AgdaSymbol{\{}\AgdaBound{n} \AgdaSymbol{:} \AgdaDatatype{ℕ}\AgdaSymbol{\}} \AgdaSymbol{→} \AgdaFunction{DNF} \AgdaSymbol{(}\AgdaInductiveConstructor{suc} \AgdaBound{n}\AgdaSymbol{)} \AgdaSymbol{→} \AgdaDatatype{Prop} \AgdaBound{n}\<%
\\
\>[6]\AgdaIndent{8}{}\<[8]%
\>[8]\AgdaField{qfree} \<[14]%
\>[14]\AgdaSymbol{:} \AgdaSymbol{\{}\AgdaBound{n} \AgdaSymbol{:} \AgdaDatatype{ℕ}\AgdaSymbol{\}} \AgdaSymbol{(}\AgdaBound{φ} \AgdaSymbol{:} \AgdaFunction{DNF} \AgdaSymbol{(}\AgdaInductiveConstructor{suc} \AgdaBound{n}\AgdaSymbol{))} \AgdaSymbol{→} \AgdaDatatype{QFree} \AgdaSymbol{(}\AgdaField{elim} \AgdaBound{φ}\AgdaSymbol{)}\<%
\\
\>[6]\AgdaIndent{8}{}\<[8]%
\>[8]\AgdaField{equiv} \<[14]%
\>[14]\AgdaSymbol{:} \AgdaSymbol{\{}\AgdaBound{n} \AgdaSymbol{:} \AgdaDatatype{ℕ}\AgdaSymbol{\}} \AgdaSymbol{(}\AgdaBound{φ} \AgdaSymbol{:} \AgdaFunction{DNF} \AgdaSymbol{(}\AgdaInductiveConstructor{suc} \AgdaBound{n}\AgdaSymbol{))} \AgdaSymbol{(}\AgdaBound{e} \AgdaSymbol{:} \AgdaDatatype{Vec} \AgdaField{Y} \AgdaBound{n}\AgdaSymbol{)} \AgdaSymbol{→} \AgdaFunction{⟦} \AgdaInductiveConstructor{E} \AgdaSymbol{(}\AgdaFunction{D.i} \AgdaBound{φ}\AgdaSymbol{)} \AgdaFunction{⟧} \AgdaBound{e} \AgdaFunction{↔} \AgdaFunction{⟦} \AgdaField{elim} \AgdaBound{φ} \AgdaFunction{⟧} \AgdaBound{e}\<%
\end{code}

Single-step elimination on conjunctions of literals (referred to here as \emph{products}) is also defined:

\begin{code}%
\>[0]\AgdaIndent{4}{}\<[4]%
\>[4]\AgdaKeyword{record} \AgdaRecord{ProdQE} \AgdaSymbol{:} \AgdaPrimitiveType{Set} \AgdaKeyword{where}\<%
\\
\>[4]\AgdaIndent{6}{}\<[6]%
\>[6]\AgdaKeyword{field}\<%
\\
\>[6]\AgdaIndent{8}{}\<[8]%
\>[8]\AgdaField{elim} \<[14]%
\>[14]\AgdaSymbol{:} \AgdaSymbol{\{}\AgdaBound{n} \AgdaSymbol{:} \AgdaDatatype{ℕ}\AgdaSymbol{\}} \AgdaSymbol{→} \AgdaFunction{Prod} \AgdaSymbol{(}\AgdaInductiveConstructor{suc} \AgdaBound{n}\AgdaSymbol{)} \AgdaSymbol{→} \AgdaDatatype{Prop} \AgdaBound{n}\<%
\\
\>[6]\AgdaIndent{8}{}\<[8]%
\>[8]\AgdaField{qfree} \<[14]%
\>[14]\AgdaSymbol{:} \AgdaSymbol{\{}\AgdaBound{n} \AgdaSymbol{:} \AgdaDatatype{ℕ}\AgdaSymbol{\}} \AgdaSymbol{(}\AgdaBound{φ} \AgdaSymbol{:} \AgdaFunction{Prod} \AgdaSymbol{(}\AgdaInductiveConstructor{suc} \AgdaBound{n}\AgdaSymbol{))} \AgdaSymbol{→} \AgdaDatatype{QFree} \AgdaSymbol{(}\AgdaField{elim} \AgdaBound{φ}\AgdaSymbol{)}\<%
\\
\>[6]\AgdaIndent{8}{}\<[8]%
\>[8]\AgdaField{equiv} \<[14]%
\>[14]\AgdaSymbol{:} \AgdaSymbol{\{}\AgdaBound{n} \AgdaSymbol{:} \AgdaDatatype{ℕ}\AgdaSymbol{\}} \AgdaSymbol{(}\AgdaBound{φ} \AgdaSymbol{:} \AgdaFunction{Prod} \AgdaSymbol{(}\AgdaInductiveConstructor{suc} \AgdaBound{n}\AgdaSymbol{))} \AgdaSymbol{(}\AgdaBound{e} \AgdaSymbol{:} \AgdaDatatype{Vec} \AgdaField{Y} \AgdaBound{n}\AgdaSymbol{)} \AgdaSymbol{→} \AgdaFunction{⟦} \AgdaInductiveConstructor{E} \AgdaSymbol{(}\AgdaFunction{P.i} \AgdaBound{φ}\AgdaSymbol{)} \AgdaFunction{⟧} \AgdaBound{e} \AgdaFunction{↔} \AgdaFunction{⟦} \AgdaField{elim} \AgdaBound{φ} \AgdaFunction{⟧} \AgdaBound{e}\<%
\end{code}

The \data{Prod} datatype is shorthand for a list of literals, and, analogously to \func{DNF}, is interpreted as a proposition using \func{P.i}.

The distribution of $\exists$ across disjunctions allows single-step elimination on products to be trivially generalized to single-step elimination on propositions in DNF (proof omitted):

\begin{code}%
\>[0]\AgdaIndent{4}{}\<[4]%
\>[4]\AgdaFunction{Prod⇒DNF.lift} \AgdaSymbol{:} \AgdaRecord{ProdQE} \AgdaSymbol{→} \AgdaRecord{DNFQE}\<%
\end{code}

The conversion of any quantifier-free proposition to DNF (\func{dnf}, \func{dnf-fwd}, \func{dnf-bwd} above) allows single-step elimination on propositions in DNF to be generalized to single-step elimination on any quantifier-free proposition (proof omitted):

\begin{code}%
\>[0]\AgdaIndent{4}{}\<[4]%
\>[4]\AgdaFunction{lift-dnf-qe} \AgdaSymbol{:} \AgdaRecord{DNFQE} \AgdaSymbol{→} \AgdaRecord{QE}\<%
\end{code}

Finally, composing these two functions allows single-step elimination on products to be generalized to single-step elimination on any quantifier-free proposition:

\begin{code}%
\>[0]\AgdaIndent{4}{}\<[4]%
\>[4]\AgdaFunction{lift-prod-qe} \AgdaSymbol{:} \AgdaRecord{ProdQE} \AgdaSymbol{→} \AgdaRecord{QE}\<%
\\
\>[0]\AgdaIndent{4}{}\<[4]%
\>[4]\AgdaFunction{lift-prod-qe} \AgdaSymbol{=} \AgdaFunction{lift-dnf-qe} \AgdaFunction{∘} \AgdaFunction{Prod⇒DNF.lift}\<%
\end{code}

The scope of the problem is therefore narrowed considerably: to obtain full quantifier elimination and decidability (recalling the results of Sections~\ref{subsec:qeqeqe} and \ref{subsec:dec}, \func{lift-qe} and \func{⟦\_⟧?} in particular), one need only create a single-step quantifier elimination procedure that acts on products.
This can be put to direct use on a number of theories, as described by Nipkow~\cite{nipkowjar}.
The following section gives an overview of one such application.

\section{The Theory of Successor}\label{sec:suc}

The example theory to which this framework is applied is the theory of successor on the natural numbers (``SN''), as presented by Herbrand~\cite{herbrand}.
Atomic formulae are equalities between terms, each of which is the application of the successor function $S$ some (known) number of times to either a variable or zero.
For example:
\[S(S(S(x))) = S(S(S(S(0))))\]
or, more concisely:
\[S^3(x) = S^4(0).\]
For convenience the previous atomic formula may be written as $x + 3 = 4$, but it is important to note that---unlike in the case of Presburger arithmetic---one may not add variables together, as this would represent an unknown number of applications of $S$.
Thus $x + 3 = y + 7$ corresponds to a valid atomic formula, while $x + y = z + 5$ does not.

The machinery developed in Section~\ref{sec:qeqe} allows the scope of quantifier elimination to be reduced to elimination of a single variable \(x\) from a conjunction of literals (equalities or negated equalities, i.e. inequalities).
For this theory, that can be accomplished via substitution; if an equality involving \(x\) is found, such as \(x + 5 = y + 3\), then substitutions are made accordingly throughout the conjunction, thereby removing \(x\), and the inequalities \(y \ne 0\) and \(y \ne 1\) are added.
On the other hand, if \(x\) only occurs in inequalities, then those inequalities may be dropped from the conjunction---each one is only false for at most one value of \(x\), so there is always some value for \(x\) which satisfies them all.\footnote{This excludes trivial inequalities such as \(x + 3 \ne x + 3\) and \(x + 2 \ne x + 4\), which are simplified instead of dropped.}

The (rather verbose) formalization of the above, omitted from this paper in the interest of brevity, results in a \record{ProdQE} object.
This is then lifted to a \record{QE} object via \func{lift-prod-qe}, leading to full quantifier elimination via \func{lift-qe} and decidability via \func{⟦\_⟧?}.

\section{Demonstration}\label{sec:demon}

The resulting decision procedure and consequences are demonstrated on several small propositions in order to give a sense of the benefits offered by a constructive approach.
The syntax for SN propositions leaves much to be desired; what they code for is indicated with comments.
First, a simple system of equalities:
\begin{code}%
\>[0]\AgdaIndent{2}{}\<[2]%
\>[2]\AgdaFunction{test₀} \AgdaSymbol{:} \AgdaDatatype{Prop} \AgdaInductiveConstructor{zero}\<%
\\
\>[0]\AgdaIndent{2}{}\<[2]%
\>[2]\AgdaFunction{test₀} \AgdaSymbol{=} \AgdaInductiveConstructor{E} \AgdaInductiveConstructor{E} \AgdaSymbol{(} \<[57]%
\>[57]\AgdaComment{-- ∃x.∃y.}\<%
\\
\>[2]\AgdaIndent{6}{}\<[6]%
\>[6]\AgdaSymbol{(}\AgdaInductiveConstructor{atom} \AgdaSymbol{(}\AgdaInductiveConstructor{S} \AgdaNumber{3} \AgdaSymbol{(}\AgdaInductiveConstructor{var} \AgdaSymbol{(}\AgdaInductiveConstructor{fsuc} \AgdaInductiveConstructor{fzero}\AgdaSymbol{))} \AgdaInductiveConstructor{==} \AgdaInductiveConstructor{S} \AgdaNumber{1} \AgdaSymbol{(}\AgdaInductiveConstructor{var} \AgdaInductiveConstructor{fzero}\AgdaSymbol{)))} \<[57]%
\>[57]\AgdaComment{-- 3 + x = 1 + y}\<%
\\
\>[0]\AgdaIndent{4}{}\<[4]%
\>[4]\AgdaInductiveConstructor{∧} \AgdaSymbol{(}\AgdaInductiveConstructor{atom} \AgdaSymbol{(}\AgdaInductiveConstructor{S} \AgdaNumber{8} \AgdaInductiveConstructor{∅} \AgdaInductiveConstructor{==} \AgdaInductiveConstructor{S} \AgdaNumber{4} \AgdaSymbol{(}\AgdaInductiveConstructor{var} \AgdaInductiveConstructor{fzero}\AgdaSymbol{)))} \<[57]%
\>[57]\AgdaComment{-- 8 = 4 + y}\<%
\\
\>[0]\AgdaIndent{4}{}\<[4]%
\>[4]\AgdaSymbol{)}\<%
\end{code}
Normalizing \func{⟦} \bound{test₀} \func{⟧?} \cons{[]} yields:\footnote{Technically, each \cons{refl} appeared as \cons{.Agda.Builtin.Equality.\_≡\_.refl}.}
\begin{code}%
\>[0]\AgdaIndent{2}{}\<[2]%
\>[2]\AgdaInductiveConstructor{yes} \AgdaSymbol{(}\AgdaNumber{2}\AgdaSymbol{,}\AgdaNumber{4}\AgdaSymbol{,}\AgdaInductiveConstructor{refl}\AgdaSymbol{,}\AgdaInductiveConstructor{refl}\AgdaSymbol{)}
\end{code}
The \num{2} and \num{4} are witnesses to the existential quantifiers, which is to say values for $x$ and $y$, and the pair of \cons{refl} constitute a proof of the inner conjunction (under the environment \asym{[}\num{4}\asym{,}\num{2}\asym{]}).

Next, a proposition with a universal quantifier is decided:
\begin{code}%
\>[0]\AgdaIndent{2}{}\<[2]%
\>[2]\AgdaComment{-- ∀x.(x=0 ∨ ∃y.x=y+1)}\<%
\\
\>[0]\AgdaIndent{2}{}\<[2]%
\>[2]\AgdaFunction{test₁} \AgdaSymbol{:} \AgdaDatatype{Prop} \AgdaInductiveConstructor{zero}\<%
\\
\>[0]\AgdaIndent{2}{}\<[2]%
\>[2]\AgdaFunction{test₁} \AgdaSymbol{=} \AgdaInductiveConstructor{A} \AgdaSymbol{((}\AgdaInductiveConstructor{atom} \AgdaSymbol{(}\AgdaInductiveConstructor{S} \AgdaNumber{0} \AgdaSymbol{(}\AgdaInductiveConstructor{var} \AgdaInductiveConstructor{fzero}\AgdaSymbol{)} \AgdaInductiveConstructor{==} \AgdaInductiveConstructor{S} \AgdaNumber{0} \AgdaInductiveConstructor{∅}\AgdaSymbol{))} \<%
\\
\>[2]\AgdaIndent{4}{}\<[4]%
\>[4]\AgdaInductiveConstructor{∨} \AgdaSymbol{(}\AgdaInductiveConstructor{E} \AgdaSymbol{(}\AgdaInductiveConstructor{atom} \AgdaSymbol{(}\AgdaInductiveConstructor{S} \AgdaNumber{0} \AgdaSymbol{(}\AgdaInductiveConstructor{var} \AgdaSymbol{(}\AgdaInductiveConstructor{fsuc} \AgdaInductiveConstructor{fzero}\AgdaSymbol{))} \AgdaInductiveConstructor{==} \AgdaInductiveConstructor{S} \AgdaNumber{1} \AgdaSymbol{(}\AgdaInductiveConstructor{var} \AgdaInductiveConstructor{fzero}\AgdaSymbol{)))))}\<%
\end{code}
\func{⟦} \bound{test₁} \func{⟧?} \cons{[]} normalizes to \cons{yes} followed by a (424-line) function that proves the inner proposition for any given $x$.

Finally, a proposition with a free variable is examined:
\begin{code}%
\>[0]\AgdaIndent{2}{}\<[2]%
\>[2]\AgdaComment{-- (x = 0) ∨ (∃y.x=y+2)}\<%
\\
\>[0]\AgdaIndent{2}{}\<[2]%
\>[2]\AgdaFunction{test₂} \AgdaSymbol{:} \AgdaDatatype{Prop} \AgdaNumber{1}\<%
\\
\>[0]\AgdaIndent{2}{}\<[2]%
\>[2]\AgdaFunction{test₂} \AgdaSymbol{=} \AgdaSymbol{((}\AgdaInductiveConstructor{atom} \AgdaSymbol{(}\AgdaInductiveConstructor{S} \AgdaNumber{0} \AgdaSymbol{(}\AgdaInductiveConstructor{var} \AgdaInductiveConstructor{fzero}\AgdaSymbol{)} \AgdaInductiveConstructor{==} \AgdaInductiveConstructor{S} \AgdaNumber{0} \AgdaInductiveConstructor{∅}\AgdaSymbol{))}  \<%
\\
\>[2]\AgdaIndent{4}{}\<[4]%
\>[4]\AgdaInductiveConstructor{∨} \AgdaSymbol{(}\AgdaInductiveConstructor{E} \AgdaSymbol{(}\AgdaInductiveConstructor{atom} \AgdaSymbol{(}\AgdaInductiveConstructor{S} \AgdaNumber{0} \AgdaSymbol{(}\AgdaInductiveConstructor{var} \AgdaSymbol{(}\AgdaInductiveConstructor{fsuc} \AgdaInductiveConstructor{fzero}\AgdaSymbol{))} \AgdaInductiveConstructor{==} \AgdaInductiveConstructor{S} \AgdaNumber{2} \AgdaSymbol{(}\AgdaInductiveConstructor{var} \AgdaInductiveConstructor{fzero}\AgdaSymbol{)))))}\<%
\end{code}
The function \func{∀-or-∃¬} (Section~\ref{subsec:dec}) is run on \func{test₂}.
As \func{test₂} is not true for all values, a counterexample is produced instead: \cons{inj₂} \asym{(} \num{1} \asym{,} \bound{...} \asym{)}, $1$ being a value for which the proposition does not hold, and \bound{...} consisting of a trivial proof that $1 \ne 0$ and a lengthy proof that no $y$ exists such that $1 = 2 + y$.

\section{Future Work}\label{sec:remarks}

While the code for the ``core'' of the theory-independent portion is relatively well-organized, the same cannot be said for the DNF-conversion portion.
The handling of trivially true or false factors in the latter is also suboptimal: those that are trivially true can (and should) be removed from products, as they have no effect other than fueling DNF explosion, and those that are false imply that the product as a whole is false, and therefore equivalent to $\bot$ (the recognition of which would greatly speed up the procedure, as well as contributing less to the aforementioned DNF explosion).

Another improvement would be the application of the framework to more expressive theories, such as Presburger arithmetic or any of the other theories discussed in Section~\ref{subsec:hist}.
In the former case, it may be possible to adapt Allais's proof~\cite{allaisgit}.

A final area of future work noted here is the implemention of \emph{reflection}, where propositions in the metatheory can be manipulated directly (or, seemingly so).
This would allow quantifier elimination to be applied to suitable Agda propositions directly, without the need to first encode them as \data{Prop}.
Such a strategy is employed in Nipkow's framework~\cite{nipkowjar} in Isabelle, and the Omega solver~\cite{omega} provides similar functionality for Presburger arithmetic in Coq (albeit as a tactic, rather than through reflection).
Agda's reflection mechanism is discussed in a more general context by van der Walt~\cite{vanderwalt2}.

\section{Acknowledgements}
This paper is adapted from the author's 2018 master's thesis in Computer Science at the University of Gothenburg.
The idea for the project came from the author's thesis supervisors Thierry Coquand and Simon Huber, with additional help and encouragement from Andreas Abel.

\newpage
\nocite{*}
\bibliographystyle{eptcs}
\bibliography{refs}
\end{document}